\documentclass[fleqn,usenatbib]{mnras}

\usepackage{newtxtext,newtxmath}
\usepackage[T1]{fontenc}

\DeclareRobustCommand{\VAN}[3]{#2}
\let\VANthebibliography\thebibliography
\def\thebibliography{\DeclareRobustCommand{\VAN}[3]{##3}\VANthebibliography}

\usepackage{graphicx}	% Including figure files

\usepackage{pdflscape}	% Landscape pages

\usepackage{tikz}

\title[Aberrations in a diffraction-limited spectrograph]{Determining the aberrations in a nearly diffraction-limited spectrograph}

\author[B. Sánchez et al.]{
B. Sánchez,$^{1}$\thanks{E-mail: beatriz@astro.unam.mx (BS)}
A. M. Watson,$^{1}$
and S. Cuevas$^{1}$
\\
$^{1}$Instituto de Astronomía, Universidad Nacional Autónoma de México, Apartado Postal 70-264, 04510 México, CDMX, Mexico\\
}

\date{Accepted XXX. Received YYY; in original form ZZZ}

\pubyear{2021}

\begin{document}
\label{firstpage}
\pagerange{\pageref{firstpage}--\pageref{lastpage}}
\maketitle

\begin{abstract}
We present a method to determine the static aberrations in a nearly diffraction-limited spectrograph introduced, for example, by alignment or manufacturing errors. We consider an instrument with two stages separated by a slit or image slicer located in the intermediate focal plane. In such a spectrograph, it is not trivial to distinguish aberrations in the first stage, before the slit, from those in the second, after the slit. However, our method achieves this. Measuring these aberrations separately opens the possibility of reducing them, by realignment or other means, and thereby improving the optical performance of the instrument. The method is based on fitting models to multiple images of a point source, with controlled displacements of the source perpendicular to the slit and controlled defocuses of the second stage or the detector. Fitting models to these images allows the determination of the aberrations in both stages. Our key discovery is that the displaced and defocused images provide additional information which allows us to break the ambiguity between the two stages. We present simulations that validate the performance of the method.

\end{abstract}


%%%%%%%%%%%%%%%%%%%%%%%%%%%%%%%%%%%%%%%%%%%%%%%%%%

%%%%%%%%%%%%%%%%% BODY OF PAPER %%%%%%%%%%%%%%%%%%

\section{Introduction}

Instruments for adaptive-optics systems are typically designed and constructed to have small optical aberrations and thereby achieve intrinsic Strehl ratios of 0.8--0.9 or more. For instruments operating in the near infrared from 1.2 to 2.5 {\micron}, this requires that the RMS aberrations be no more than about 100 nm ($\lambda/13$ to $\lambda/20$). This can be achieved either by manufacturing and aligning an instrument to high precision or by using the adaptive-optic system to compensate the instrument's aberrations (a process known as “non-common-path error correction”). Both approaches require a means to determine the aberrations in the instrument.

The aberrations in imagers can be determined by well-known methods such as Shack-Hartmann \citep{1978ost..conf.....M}, curvature \citep{1988ApOpt..27.1223R}, phase diversity \citep{2003A&A...399..373B,2003A&A...399..385H}, and Donut \citep{2006PASP..118.1165T}.

On the other hand, spectrographs often have two stages with a slit or image slicer in the intermediate focal plane. In some, one can estimate the total aberration of both stages by removing the slit or using a wide slit and using the techniques appropriate for imagers. However, there are potentially two problems with this approach. First, in some instruments it is not possible (or at least not feasible) to remove the slit or use a wide slit. Second, this approach gives the sum of the aberrations in the two stages but cannot distinguish, for example, an aberration present in the one stage from an equal one in the other. A pathological case would be significant aberrations present in both stages with equal magnitude but opposite signs. In this case, the total aberration would be zero, but the performance of the spectrograph could be severely degraded as the image projected on the slit or image slicer would be aberrated and not limited by diffraction.

With these motivations, in this work we present a method to measure the aberrations in a spectrograph that is capable of separating the aberrations in one stage from those in the other.

Our method is based on acquiring images of a point source with the science detector, first centered on the slit and then with controlled small displacements perpendicular to the slit. At each slit position, images are taken both in focus and with a controlled defocus in the second stage or of the detector. This set of images provides the information required to allow us to determine the aberrations in each stage of the spectrograph without ambiguity.

In a space-based telescope, the point source could be an astronomical source. The first stage of the spectrograph would then encompass the telescope and all of the optics before the slit. In a ground-based telescope, atmospheric turbulence means that we cannot realistically use astronomical sources, since even with an adaptive-optics system the wave-front delivered to the instrument typically has time-dependent aberrations. Instead, we require either a telescope simulator or a means to place a point source in an intermediate focal plane before the slit. The first stage then encompasses only the optics after the position of the point source, which will often exclude the telescope. Depending on the application, this can be an advantage or a disadvantage.

Our method requires no unusual auxiliary equipment, just the means to place a point source in a focal plane prior to the slit (or to feed the spectrograph with a telescope simulator), to make controlled displacements of this point source perpendicular to the slit (or the slit with respect to the point source), and to defocus the second stage or move the detector.

Our immediate motivation for this work was to develop a means to estimate the aberrations in the FRIDA instrument \citep{2016SPIE.9908E..0PW}. FRIDA will be a main science instrument for the GTCAO adaptive-optics system of the GTC telescope \citep{Devaney-2000,2016SPIE.9908E..0PW,2019hsax.conf..536B}. The input focal plane of FRIDA coincides with the output focal plane of GTCAO. FRIDA will work in the near infrared and provide both imaging and integral-field spectroscopy using an image slicer with 30 slices. FRIDA will be outside of the control loop of the adaptive optics system, so other techniques must be used to determine the internal aberrations.

In imaging mode, FRIDA conceptually has a single collimator-camera stage, as a fold-mirror is used to bypass the spectrograph and send light directly to the detector. In this mode, the aberrations can be determined using the methods mentioned earlier for imagers. The FRIDA science detector is mounted on a linear focus stage, which allows us to easily take the out-of-focus images required by these methods.

On the other hand, in integral-field spectroscopy mode, the fold mirror is removed and the instrument has two collimator-camera stages with a slit in the intermediate focal plane. More precisely, the image slicer is in the intermediate focal plane, and since each slice has its own optics, effectively we have 30 semi-independent spectrographs, each with slightly different optics and slightly different aberrations. Furthermore, the image slicer cannot be removed from the optical path of the spectrograph, although the spectrograph gratings can be replaced with a mirror. Therefore, we will use the method developed here to determine the internal aberrations of each optical path through the spectrograph by imaging through each of the slits. We will do so by placing an illuminated point source in the input focal plane of FRIDA at the position corresponding to each slit. This is facilitated by the presence of a mechanism to change field and calibration masks in the input focal plane. Our method will allow us to separately determine the aberrations in the common first collimator-camera stage and the semi-independent second collimator-camera stages. 

Our method is suitable for estimating the static aberrations in the instrument. It is not suitable for real-time correction of atmospheric aberrations, as it uses the science detector, which has a relatively high read noise, and it requires controlled motions of both the object and the science detector, which will interfere with science exposures.

Despite our focus on FRIDA, our method is applicable to other similar instruments that aim to be limited by diffraction, that have long slits or image slicers, and that require characterization of their optics. Examples of instruments of this nature are SINFONI \citep{2003SPIE.4841.1548E}, CRIRES \citep{2004SPIE.5492.1218K}, and MUSE \citep{2010SPIE.7735E..08B}.

This article is organized as follows. In section~\ref{section:model} we summarize the theory of the formation of images in a spectrograph with two stages and an intermediate slit. This is the basis for all of the work presented here and establishes our notation. In section~\ref{section:slit} we show simulations that demonstrate the effect of aberrations in both stages of the spectrograph on the final image. An understanding of the consequences of the aberration and in particular certain ambiguities is helpful to understand the algorithm. In section~\ref{section:algorithm} we describe our algorithm. In section~\ref{section:tests} we present tests of our algorithm. In section~\ref{section:future-plans} and \ref{section:conclusions}, we indicate possible lines of future work and discuss our conclusions. In appendix~\ref{appendix:catalog} we show a catalog of aberrated images to complement the discussion in section~\ref{section:slit}.

\section{Modelling image formation}
\label{section:model}

\begin{figure}
    \centering
    \includegraphics[width=\linewidth]{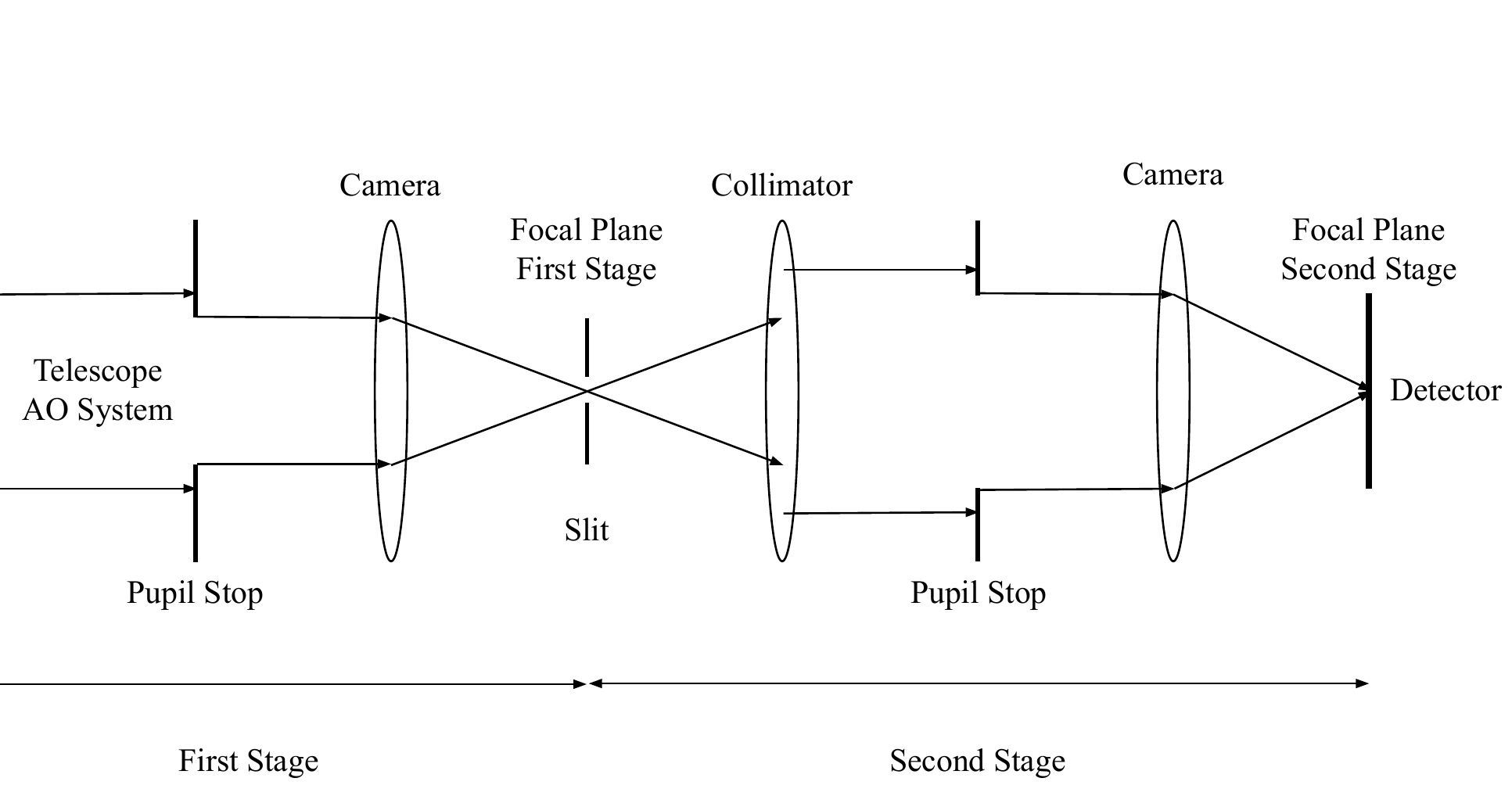}
    \caption{A spectrograph with two stages and an intermediate slit.}
    \label{figure:layout}
\end{figure}

Figure~\ref{figure:layout} shows schematically the layout of an optical system with two stages and a slit in the intermediate focal plane. This is the type of optical system with which we will be concerned in this work.

In an instrument limited by seeing, the size of the pupil in the second stage is typically only slightly larger than that in the first stage. However, in an instrument limited by diffraction, with a slit width close to the full-width at half-maximum (FWHM) of the point-spread function (PSF), the size of the pupil in the second stage is often significantly larger to avoid diffraction losses. Throughout this work, we will take the pupil in the second stage to be twice the diameter of the pupil in the first stage, although this choice is not essential for our method.

In the system under consideration, there is no dispersive element (grating, prism, or grism) in the second stage. It is not, in fact, a spectrograph but rather a system that images through a slit. Most real instruments will indeed have dispersive elements in the second stage. We offer three suggestions for eliminating their effects. First, it might be possible to replace gratings with mirrors or remove prisms/grisms. This is possible in FRIDA, which has two mirrors in the grating turret originally intended to aid precise acquisition through the image slicer. Second, it might be possible to use or rotate the elements into zeroth order. Finally, it might be possible to mitigate the dispersive element using monochromatic light. If none of these are feasible, it might be possible to extend the method presented here to work with dispersed light, but we suspect it will lose sensitivity to aberrations perpendicular to the slit.

Our physical polar coordinates in pupil or focal planes are $\mathbfit{x} = (\rho ,\theta)$. We define $\theta = 0$ to be parallel to the slit. We define the radii of the pupils in the first and second stages to be $R_1$ and $R_2$. We define normalized radial coordinates $\rho_1 \equiv \rho / R_1$ and $\rho_2 \equiv \rho / R_2$ along with the corresponding normalized polar coordinates $\mathbfit{x}_1 \equiv (\rho_1,\theta)$ and $\mathbfit{x}_2 \equiv (\rho_2,\theta)$.

We define $P_1(\mathbfit{x}_1)$ and $P_2(\mathbfit{x}_2)$ to be the pupil transmission functions, which are 1 where the pupil transmits light and 0 where it blocks it. In this work, we will deal exclusively with circular pupils without central obstructions, and so have
\begin{align}
    P_1(\mathbfit{x}_1) = \left\{\begin{array}{ll}1&\mbox{for $\rho_1 \le 1$, and}\\0&\mbox{for $\rho_1 > 1$,}\end{array}\right.
\end{align}
and a similar definition for $P_2$. 

We define the wave-front errors in the first and second stages to be $W_1(\mathbfit{x}_1)$ and $W_2(\mathbfit{x}_2)$. We represent these as sums of Zernike polynomials $Z_j$,
\begin{align}
    W_1(\mathbfit{x}_1) &= \sum_{j=1}^{N_\mathrm{Z}} a_j Z_j(\mathbfit{x}_1)\\
    \intertext{and}
    W_2(\mathbfit{x}_2) &= \sum_{j=1}^{N_\mathrm{Z}} b_j Z_j(\mathbfit{x}_2).
\end{align}
As we use the \cite{1976JOSA...66..207N} normalization, the coefficients $a_j$ and $b_j$ are the RMS amplitudes of the aberrations corresponding to each Zernike polynomial. As a convenience to the reader, in Table~\ref{table:zernikes} we give the first 11 Zernike polynomials, which correspond to tilts, defocus, and the classical third-order aberrations.

\begin{table}[]
    \centering
    \caption{The first 11 Zernike polynomials from \protect\cite{1976JOSA...66..207N}.}
    \label{table:zernikes}
    \begin{tabular}{lll}
        \hline
        $j$&$Z_j$&Name  \\
        \hline
        1&1&Piston\\
        2&$4^{1/2}(\rho) \cos\theta$&Tilt in $y$\\
        3&$4^{1/2}(\rho) \sin\theta$&Tilt in $x$\\
        4&$3^{1/2}(2\rho^2-1)$&Defocus\\
        5&$6^{1/2}(\rho^2) \sin 2\theta$&Astigmatism at 0 deg\\
        6&$6^{1/2}(\rho^2) \cos 2\theta$&Astigmatism at 45 deg\\
        7&$8^{1/2}(3\rho^3 -2\rho)\sin \theta$&Coma at 90 deg\\
        8&$8^{1/2}(3\rho^3 -2\rho)\cos \theta$&Coma at 0 deg\\
        9&$8^{1/2}(\rho^3)\sin 3\theta$&Trefoil at 30 deg\\
        10&$8^{1/2}(\rho^3)\cos 3\theta$&Trefoil at 0 deg\\
        11&$5^{1/2}(6\rho^4-6\rho^2+1)$&Spherical\\
        \hline
    \end{tabular}
\end{table}

We define $S(\mathbfit{x})$ to be the slit transmission function, which is 1 where the slit transmits light and 0 where it blocks it. 

%With these definitions, we can write the intensity of the image of a point source formed by a two-stage system with an intermediate slit as
%\begin{align}
%I_2(\mathbfit{x}) \propto \left|F\left(P_2(\mathbfit{x}_2)e^{i\phi_2(\mathbfit{x}_2)}F^{-1}\left[S(\mathbfit{x})F\left(P_1(\mathbfit{x}_1)e^{i\phi_1(\mathbfit{x}_1)}\right)\right]\right)\right|^2,
%\label{eq:I}
%\end{align}
%in which $\phi_1(\mathbfit{x}_1) \equiv 2\pi W_1(\mathbfit{x}_1)/\lambda$, $\phi_2(\mathbfit{x}_1) \equiv 2\pi W_2(\mathbfit{x}_2)/\lambda$, and we denote by $F$ and $F^{-1}$ the Fourier transform and inverse Fourier transform operators. The PSF is then given by $I/\max(I)$.

With these definitions, we can write the phases $\phi_1$ and $\phi_2$ in the two pupil planes, the amplitudes $A_1$ and $A_2$ in the two pupil planes, and the intensities $I_1$ and $I_2$ in the two focal planes as
\begin{align}
    \phi_1 &= 2\pi W_1(\mathbfit{x}_1)/\lambda,\\
    A_1 &= P_1(\mathbfit{x}_1),\\
    I_1 &= F\left(A_1(\mathbfit{x}_1)e^{i\phi_1(\mathbfit{x}_1)}\right),\\
    \phi_2 &= P_2 \arg\left(F^{-1}\left(SI_1\right)\right) + 2\pi W_2(\mathbfit{x}_2)/\lambda,\\
    A_2 &= P_2 \bmod\left(F^{-1}\left(SI_1\right)\right)\mbox{, and}\\
I_2 &= F\left(A_2(\mathbfit{x}_2)e^{i\phi_2(\mathbfit{x}_2)}\right),
\end{align}
in which $F$ and $F^{-1}$ are appropriately-normalized Fourier transfer and inverse Fourier transfer operators. The final PSF is then given by $I_2/\max(I_2)$.

Numerically, we simulate the system using discrete Cartesian grids in the focal and pupil planes. One consequence of this is that we simulate a detector with finite-sized pixels.

\section{Aberrations combined with a slit}
\label{section:slit}

In this section, we use the above equations to simulate the optical system with the aim of describing and understanding the impact on the final image of aberrations both before and after the slit and by the masking by the slit of the aberrations in the first stage. This is important for understanding our subsequent results.

The systems we have simulated are almost diffraction-limited, with aberrations sufficiently small in magnitude that the Strehl ratio is always at least 0.85. We assumed that the pixel size is $\lambda/3D$ and that the slit width is $\lambda/D$ (3 pixels), which is approximately the FWHM of the diffraction-limited PSF. We consider low-order aberrations, up to and including spherical aberration (see Table~\ref{table:zernikes}), without loss of generality.

We show results at different points in the optical system. See, for example, Figure~\ref{figure:example-slit}. From left to right, the columns show: the phase $\phi_1$ in the pupil plane of the first stage; the image $I_1$ in the exit focal plane of the first stage just before the slit; the image $SI_1$ in the exit focal plane of the first stage just after the slit (which is vertical); the amplitude $A_2$  of the intensity in the pupil plane of the second stage; the phase $\phi_2$ of the intensity in the pupil plane of the second stage; the image  $I_2$ in the exit focal plane of the second stage; a profile $I_2(y=0)$ through the PSF perpendicular to the slit; and finally a profile $I_2(x=0)$  through the PSF parallel to the slit.

\subsection{Effects of the slit and tilts}

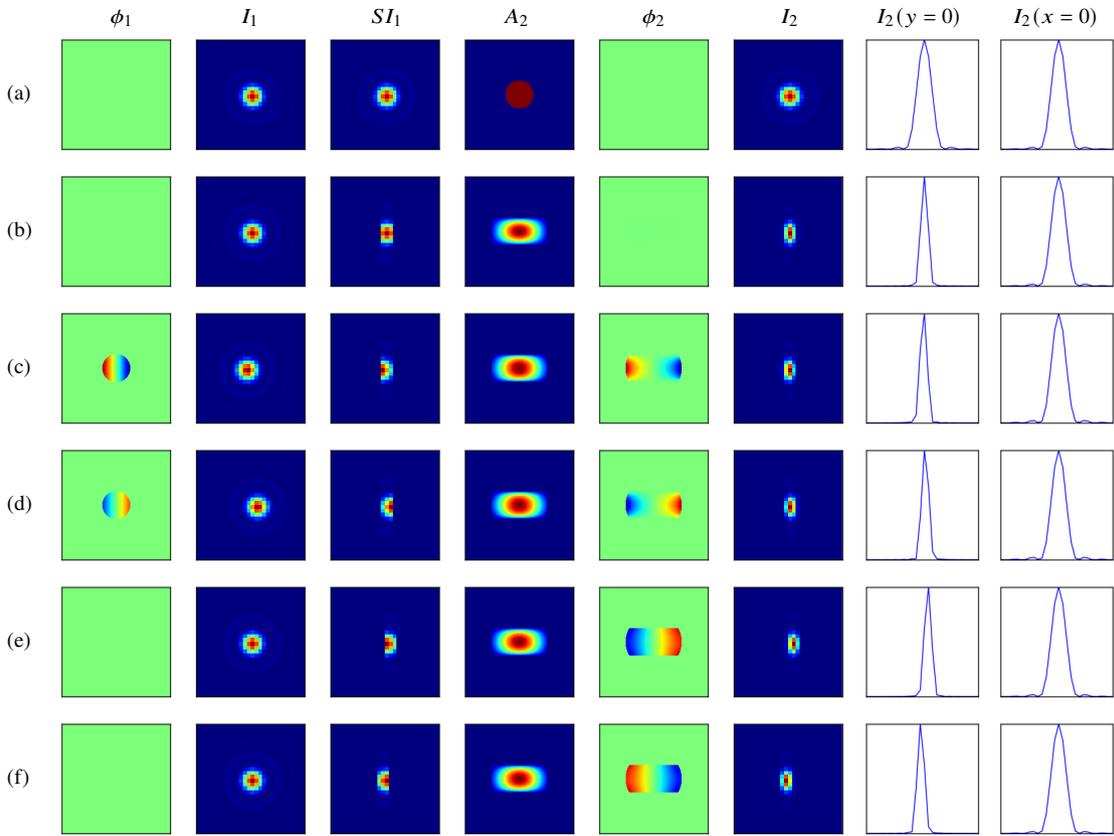
\begin{figure*}
    \centering
    \begin{tikzpicture}
    \node (a) at (0,-1) {\includegraphics[width=\linewidth]{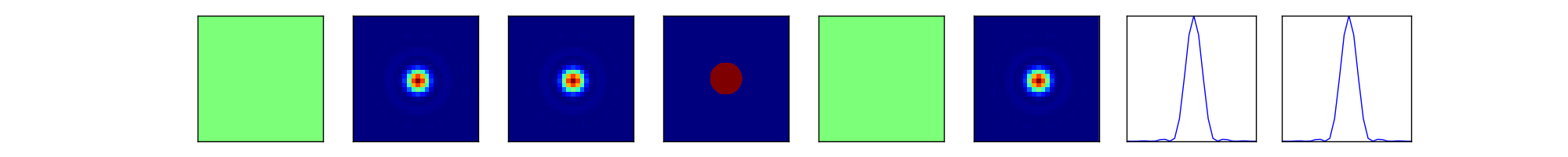}};
    \node (b) at (0,-2.8) {\includegraphics[width=\linewidth]{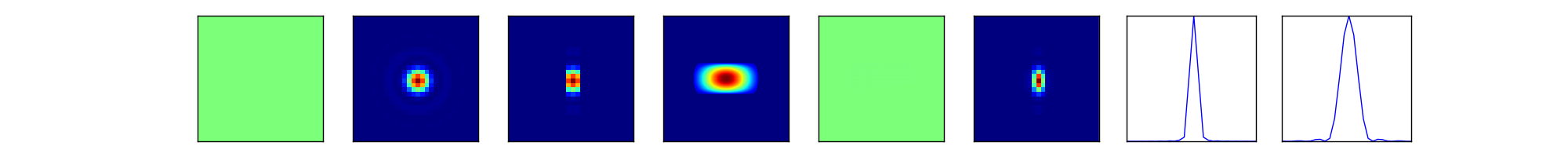}};
    \node (c) at (0,-4.6)  {\includegraphics[width=\linewidth]{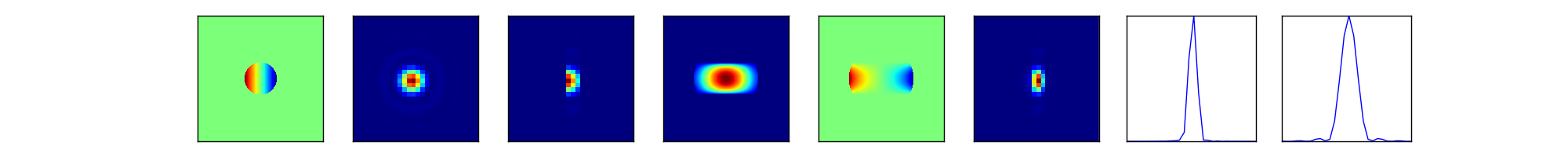}};
    \node (d) at (0,-6.4)  {\includegraphics[width=\linewidth]{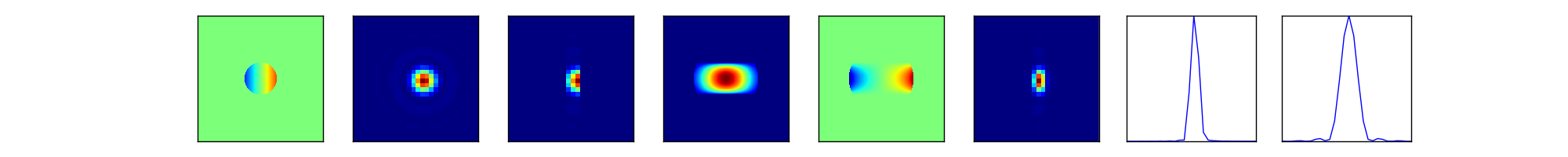}};
    \node (e) at (0,-8.2)  {\includegraphics[width=\linewidth]{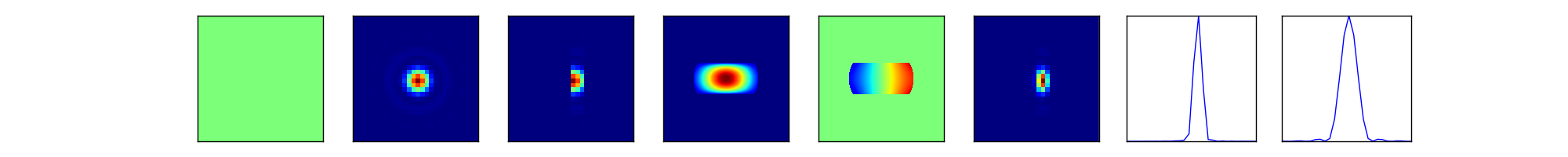}};
    \node (f) at (0,-10.0)  {\includegraphics[width=\linewidth]{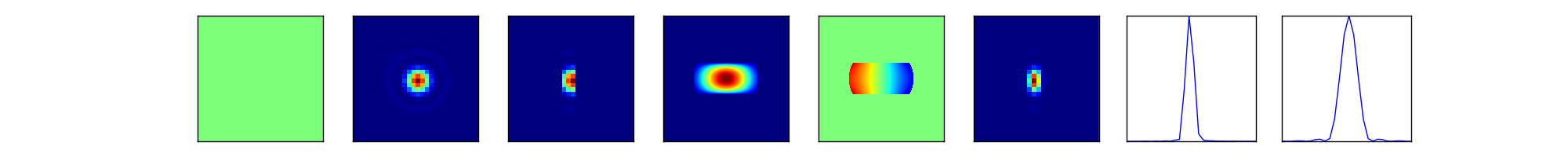}};
    \node at (-5.9,0) [anchor=mid] {$\phi_1$};
    \node at (-4.2,0) [anchor=mid] {$I_1$};
    \node at (-2.4,0) [anchor=mid] {$SI_1$};
    \node at (-0.7,0) [anchor=mid] {$A_2$};
    \node at (+1.1,0) [anchor=mid] {$\phi_2$};
    \node at (+2.9,0) [anchor=mid] {$I_2$};
    \node at (+4.6,0) [anchor=mid] {$I_2 (y = 0)$};
    \node at (+6.4,0) [anchor=mid] {$I_2 (x = 0)$};
    \node at (-7.5,-1) [anchor=west] {(a)};
    \node at (-7.5,-2.8) [anchor=west] {(b)};
    \node at (-7.5,-4.6) [anchor=west] {(c)};
    \node at (-7.5,-6.4) [anchor=west] {(d)};
    \node at (-7.5,-8.2) [anchor=west] {(e)};
    \node at (-7.5,-10.0) [anchor=west] {(f)};
    \end{tikzpicture}
    \caption{Six sequences of images showing a point source imaged by a perfect optical system. The pixel size in the focal planes is $\lambda/3D$. From left to right, the panels show: the phase $\phi_1$ in the pupil plane of the first stage; the image $I_1$ in the exit focal plane of the first stage just before the slit; the image $SI_1$ in the exit focal plane of the first stage just after the slit (which is vertical); the amplitude $A_2$ of the intensity in the pupil plane of the second stage; the phase $\phi_2$ of the intensity in the pupil plane of the second stage; the image $I_2$ in the exit focal plane of the second stage; a profile $I_2(y=0)$ through the PSF in the exit focal plane perpendicular to the slit; and finally a profile $I_2(x=0)$ through the PSF parallel to the slit. In row (a), the system has no slit. In the others, the system has a slit of width of $\lambda/D$ (3 pixels), and we observe the effects of diffraction as a broadening of the image in the pupil plane of the second stage. In row (b), the slit is centered and no effect is seen on the phase in the pupil plane of the second stage. However, in rows (c) and (d), the slit is centered, but the image is displaced by $\lambda/3D$ (1 pixel) to the left and right. In these cases, we see non-uniform gradients in the phase in the second pupil plane because of the asymmetric illumination of the slit.
    In rows (e) and (f), the image is centered, but the slit is displaced by $\lambda/3D$ (1 pixel) to the right and left.
    Again, we see gradients in the phase in the second pupil plane, now a combination of a uniform gradient from the displacement of the slit and the previous non-uniform gradient from the asymmetric illumination of the slit. In the profiles of the PSF perpendicular to the slit, we see that the slit and second stage eliminate much of the information in the PSF, especially in the wings.}
    \label{figure:example-slit}
    \label{figure:panels}
\end{figure*}

We consider initially a system without aberrations and apply only tilts and displacements of the slit. In Figure~\ref{figure:example-slit} we show six sequences of images from different configurations. By comparing these, we can clearly discern the impact of the slit.

In row (a), the slit is absent, and we see that the intensity in the pupil plane of the second stage is uniform and equal in size to the intensity in the pupil plane of the first stage. However, when we consider the other sequences which do have slits, we see that diffraction from the slit causes the intensity in the pupil of the second stage to be broadened perpendicular to the slit and to be non-uniform.

In row (b), the slit is centered, and the phase in the second pupil plane is constant. However, in rows (c) and (d), the images of the point source on the slit are displaced by $\lambda/3D$ (1 pixel) to the right and left. In these cases, we see non-uniform gradients in the phase in the second pupil plane because of the asymmetric illumination of the slit. In the core, where the illumination is more asymmetric, the gradient is stronger.

In rows (e) and (f), the image of the point source is centered, but the slit is displaced by $\lambda/3D$ to the left and right. Again, we see gradients in the phase in the second pupil plane, now a combination of a uniform gradient from the displacement of the slit and the previous non-uniform gradient from the asymmetric illumination of the slit. 

Looking at the final focal plane, we see two important effects of the slit. First, the slit removes information in the wings of the PSF. Second, the final image is more compact perpendicular to the slit.

\subsection{Effects of aberrations}

We now consider the effect of aberrations in isolation. That is, we only apply one Zernike term to the pupil of either the first or second stages.

\begin{figure*}
    \centering
    \begin{tikzpicture}
    \node (a) at (0,-1) {\includegraphics[width=\linewidth]{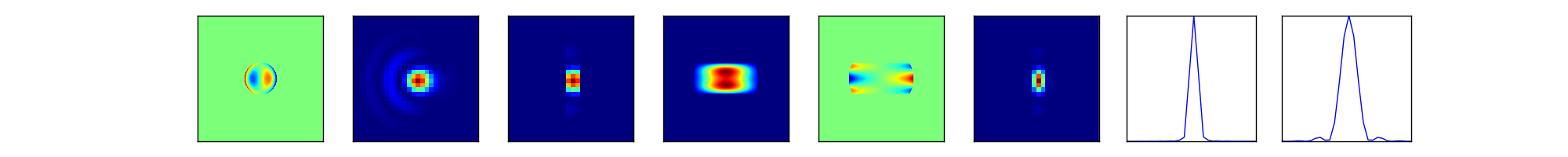}};
    \node (b) at (0,-2.8) {\includegraphics[width=\linewidth]{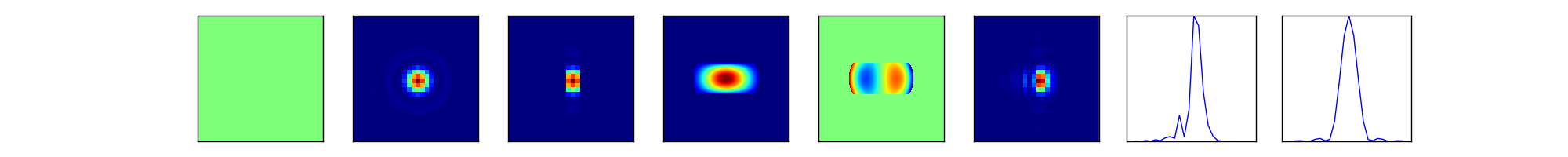}};
    \node at (-5.9,0) [anchor=mid] {$\phi_1$};
    \node at (-4.2,0) [anchor=mid] {$I_1$};
    \node at (-2.4,0) [anchor=mid] {$SI_1$};
    \node at (-0.7,0) [anchor=mid] {$A_2$};
    \node at (+1.1,0) [anchor=mid] {$\phi_2$};
    \node at (+2.9,0) [anchor=mid] {$I_2$};
    \node at (+4.6,0) [anchor=mid] {$I_2 (y = 0)$};
    \node at (+6.4,0) [anchor=mid] {$I_2 (x = 0)$};
    \node at (-8.7,-1) [anchor=west] {(a) $a_7 \approx \lambda/12$};
    \node at (-8.7,-2.8) [anchor=west] {(b) $b_7 \approx \lambda/12$};
    \end{tikzpicture}
    \caption{The effect of coma perpendicular to the slit. In the row (a) the system has a coma perpendicular to the slit in the first stage with $a_7 \approx \lambda/12$. In row (b), the coma is in the second stage with $b_7 \approx \lambda/12$. In both cases, the slit is centered. Comparing the final images in both cases, we see that in the first case the slit largely removes the asymmetry in the wings of the PSF. See Figure~\ref{figure:panels} for a description of the panels in each row. The pixel size of $3\lambda/D$ and the slit width is $\lambda/D$ (3 pixels).}
    \label{figure:example-x-coma}
\end{figure*}
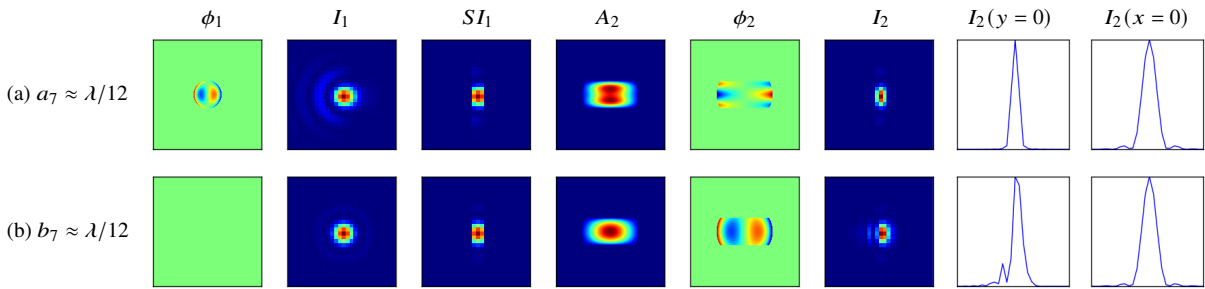

In Figure~\ref{figure:example-x-coma} we see the effect of coma perpendicular to the slit in both stages. In row (a) the system has a coma perpendicular to the slit in the first stage with $a_7 \approx \lambda/12$. In row (b), the coma is in the second stage with $b_7\approx \lambda/12$. In both cases, the slit is centered and has a width of $\lambda/D$ (3 pixels). Considering the final images, we see that coma in the second stage leaves the final image with the characteristic asymmetry associated with coma in a classical system. However, coma in first stage gives a largely symmetric final image. The effect of the slit is clear. The coma in the first stage causes the image before the slit to have a large asymmetry in the wings, but the slit masks these asymmetries and only allows the largely symmetric core to pass into the second stage. This simple example shows the complexity of the interaction of the slit and aberrations.

In Appendix~\ref{appendix:catalog}, we show similar examples of all aberrations $a_3$ to $a_{11}$ and $b_3$ to $b_{11}$. While the difference between aberrations in the first and second stages is perhaps clearest in the case of coma perpendicular to the slit, similar effects are seen in other aberrations. 

\section{Algorithm}
\label{section:algorithm}

In this section we present our algorithm for determining the aberrations in a two-stage optical system with a slit in the intermediate focal plane. The algorithm is able to determine the aberrations in the first and second stages separately.

We begin with a few words on the development of the algorithm. We worked with simulated images (with simulated noise) of point sources to be able to evaluate different approaches to fitting models to these images. We started by fitting models to a single in-focus simulated image centered on the slit, solving for the aberration coefficients $a_3$ to $a_{11}$ and $b_3$ to $b_{11}$. As might be expected, this was not very successful. One problem is that slits as narrow as $\lambda/D$ eliminate the wings of the aberrated PSF formed by the first stage, specifically the wings perpendicular to the slit. This caused large errors and  ambiguities or aliasing in the values of the aberration coefficients. The second is that there was an ambiguity in the sign of every aberration coefficient, as the final images are identical when the sign of each aberration coefficient is flipped.

To improve on this situation, we extended the algorithm to include more information. We now fit a total of six images of point sources. The first three images are in-focus, one centered on the slit and two with equal-but-opposite controlled displacements perpendicular to the slit. These images give more complete information on the parts of the wings of the PSF which previously were masked by the slit. The other three images are out-of-focus, with a controlled defocus applied in the second stage, and again, one is centered and two have the same displacements as previously. These images break the ambiguity in the sign of the aberration coefficients. Instead of displacing the point source, one could equally displace the slit.

In FRIDA, we will apply the controlled displacements by using small rotations of the focal-plane mask mechanism holding the point-source mask in the entrance focal plane and applying the controlled defocus with small displacements of the focus mechanism that moves the detector. In other instruments, other means might be more appropriate.

Our algorithm has the following steps:

\begin{enumerate}
    \item[1.] Acquire the 6 images of a point source. We denote the pixel values in these images as $I_{2i}$, in which the pixel index $i$ ranges from 1 to $N$. In practice, we select appropriate windows around the final image to reduce the number of pixels under consideration.
    \begin{enumerate}
        \item Acquire 3 in-focus images. One with the  source centered in the slit and two with equal-but-opposite controlled displacements perpendicular to the slit.
        \item Apply a controlled defocus $db_4$ to the second stage. 
        \item Acquire 3 out-of-focus images. One with the  source centered in the slit and two with equal-but-opposite controlled displacements perpendicular to the slit.
    \end{enumerate}
    \item[2.] Generate the initial parameters: the aberration coefficients $a_2$ to $a_{11}$ and $b_2$ to $b_{11}$, the normalization of the signal in each image $c_1$ to $c_6$, and the background level $b$ in the images. We typically chose aberration coefficients that are close to diffraction-limited.
    \item[3.] Generate 6 model images using the parameters. We denote the pixel values in these images as $M_{2i}$, in which the pixel index $i$ ranges from 1 to $N$. In practice, we select appropriate windows around the final image to reduce the number of pixels under consideration.
    \item[4.] Calculate the value of
    $\bar\chi^2$ according to 
    \begin{equation}
        \bar\chi^2 = \frac{1}{N} \sum_{i=1}^N \frac{g^2(I_{2i} - M_{2i})^2}{g(I_{2i} - b) + \sigma_\mathrm{b}^2},
    \end{equation}
    in which $g$ is the gain in electrons per ADU and $\sigma_\mathrm{b}$ is the noise in the background in electrons.
    \item[5.] Repeat stages 3 and 4, while varying the parameters to minimize $\bar\chi^2$. In practice, we use the Nelder-Mead optimization method, also know as the downhill-simplex or amoeba method, and terminate when the changes in $\bar\chi^2$ are sufficiently small.
    \item[6.] Repeat stages 3 to 5, but calculating the value of
    $\bar\chi^2$ according to 
    \begin{equation}
        \bar\chi^2 = \frac{1}{N} \sum_{i=1}^N \frac{g^2(I_{2i} - M_{2i})^2}{g(M_{2i} - b) + \sigma_\mathrm{b}^2}.
    \end{equation}
\end{enumerate}

Notice that we iterate twice, once using the simulated data to estimate the noise in the calculation of $\bar\chi^2$ and then again using the model. We find that the first iteration gives better convergence, while the second avoids the well-known bias.

Our experience suggests that for images with relatively high Strehl ratios, the point source should be displaced by 0.5 to 1.0 times the width of the slit and the controlled defocus should be $\lambda/10$ to $\lambda/2$ RMS. We typically use 2/3 of the slit width and $\lambda/7$ RMS defocus.

The statistical uncertainties in the aberration coefficients can be estimated either empirically from multiple images or by boot-strapping them by applying the algorithm to multiple simulated images with the same aberration coefficients as those determined for the real image and similar noise properties.

The tilts corresponding to displacements parallel to the slit, $a_2$ and $b_2$, produce identical displacements in the final focal plane and our algorithm is not able to distinguish them. Furthermore, the total tilt parallel to the slit $a_2 + b_2$ is not interesting optically, since it is degenerate with the position of the illuminating point source and the detector. For these reasons, from this point onwards we will ignore $a_2$ and $b_2$. In our code, to artificially break this degeneracy, we hold $a_2$ at 0.

\section{Tests}
\label{section:tests}

\subsection{Single aberrations}

\begin{table*}
    \centering
    \caption{Tests with single aberrations $a_3$ to $a_{11}$ in the first stage.}
    \label{table:test-single-a}
    \begin{tabular}{lcccccccccc}
    \hline
    &$a_3$&$a_4$&$a_5$&$a_6$&$a_7$&$a_8$&$a_9$&$a_{10}$&$a_{11}$\\
    \hline
$\hat a_j$&1.002&1.004&1.001&1.009&1.000&1.007&0.999&1.006&1.015\\
$\hat\sigma$&0.001&0.002&0.000&0.001&0.001&0.000&0.002&0.001&0.000\\
\hline
$\hat a_{ 3}$&       &$+0.00$&$+0.00$&$+0.00$&$+0.00$&$+0.00$&$+0.00$&$+0.00$&$+0.00$\\
$\hat a_{ 4}$&$+0.01$&       &$+0.00$&$+0.00$&$+0.01$&$+0.00$&$+0.01$&$+0.01$&$+0.01$\\
$\hat a_{ 5}$&$+0.00$&$+0.00$&       &$+0.00$&$+0.00$&$+0.00$&$+0.00$&$+0.00$&$+0.00$\\
$\hat a_{ 6}$&$+0.01$&$+0.01$&$+0.01$&       &$+0.01$&$+0.01$&$+0.01$&$+0.01$&$+0.00$\\
$\hat a_{ 7}$&$+0.00$&$+0.00$&$+0.00$&$+0.00$&       &$+0.00$&$+0.00$&$+0.00$&$+0.00$\\
$\hat a_{ 8}$&$+0.00$&$+0.00$&$+0.00$&$+0.00$&$+0.00$&       &$+0.00$&$+0.00$&$+0.00$\\
$\hat a_{ 9}$&$+0.00$&$+0.00$&$+0.00$&$+0.00$&$+0.00$&$+0.00$&       &$+0.00$&$+0.00$\\
$\hat a_{10}$&$+0.00$&$+0.00$&$+0.00$&$+0.00$&$+0.00$&$+0.00$&$+0.00$&       &$+0.00$\\
$\hat a_{11}$&$+0.00$&$+0.00$&$+0.01$&$+0.01$&$+0.00$&$+0.00$&$+0.01$&$+0.00$&       \\
\hline
$\hat b_{ 3}$&$+0.00$&$+0.00$&$+0.00$&$+0.00$&$+0.00$&$+0.00$&$+0.00$&$+0.00$&$+0.00$\\
$\hat b_{ 4}$&$+0.00$&$+0.01$&$-0.01$&$+0.01$&$+0.01$&$+0.00$&$+0.01$&$+0.01$&$+0.00$\\
$\hat b_{ 5}$&$+0.00$&$+0.00$&$+0.00$&$+0.00$&$+0.00$&$+0.00$&$+0.00$&$+0.00$&$+0.00$\\
$\hat b_{ 6}$&$-0.01$&$-0.01$&$-0.03$&$+0.00$&$+0.00$&$-0.01$&$-0.01$&$+0.00$&$-0.01$\\
$\hat b_{ 7}$&$+0.00$&$+0.00$&$+0.00$&$+0.00$&$+0.00$&$+0.00$&$+0.00$&$+0.00$&$+0.00$\\
$\hat b_{ 8}$&$+0.00$&$+0.01$&$-0.01$&$+0.00$&$-0.01$&$-0.01$&$+0.00$&$+0.01$&$+0.00$\\
$\hat b_{ 9}$&$+0.00$&$+0.00$&$+0.00$&$+0.00$&$+0.00$&$+0.00$&$+0.00$&$+0.00$&$+0.00$\\
$\hat b_{10}$&$-0.01$&$+0.01$&$-0.01$&$+0.00$&$-0.01$&$-0.01$&$+0.00$&$+0.01$&$+0.00$\\
$\hat b_{11}$&$+0.00$&$+0.00$&$+0.00$&$+0.00$&$+0.00$&$+0.00$&$+0.00$&$+0.00$&$+0.00$\\
    \hline
    \end{tabular}
\end{table*}

\begin{table*}
    \centering
    \caption{Tests with single aberrations $b_3$ to $b_{11}$ in the second stage.}
    \label{table:test-single-b}
    \begin{tabular}{lcccccccccc}
    \hline
    &$b_3$&$b_4$&$b_5$&$b_6$&$b_7$&$b_8$&$b_9$&$b_{10}$&$b_{11}$\\
    \hline
$\hat b_j$&1.002&1.017&1.006&0.879&1.007&0.990&1.016&1.001&1.013\\
$\hat \sigma$&0.002&0.002&0.000&0.005&0.000&0.006&0.000&0.001&0.000\\
\hline
$\hat a_{ 3}$&$+0.00$&$+0.00$&$+0.00$&$+0.01$&$+0.00$&$-0.01$&$-0.01$&$+0.00$&$+0.00$\\
$\hat a_{ 4}$&$+0.00$&$+0.01$&$+0.00$&$+0.04$&$+0.01$&$+0.01$&$+0.00$&$+0.00$&$-0.01$\\
$\hat a_{ 5}$&$+0.00$&$+0.00$&$+0.00$&$+0.00$&$+0.00$&$+0.00$&$+0.00$&$+0.00$&$+0.00$\\
$\hat a_{ 6}$&$+0.01$&$+0.01$&$+0.01$&$+0.05$&$+0.01$&$+0.01$&$+0.01$&$+0.00$&$+0.00$\\
$\hat a_{ 7}$&$-0.01$&$+0.00$&$-0.01$&$+0.01$&$-0.01$&$-0.01$&$-0.01$&$+0.00$&$+0.00$\\
$\hat a_{ 8}$&$+0.00$&$+0.00$&$+0.00$&$+0.00$&$+0.00$&$+0.00$&$+0.00$&$+0.00$&$+0.00$\\
$\hat a_{ 9}$&$+0.00$&$+0.00$&$+0.00$&$+0.00$&$+0.00$&$+0.00$&$+0.00$&$+0.00$&$+0.00$\\
$\hat a_{10}$&$+0.00$&$+0.00$&$+0.00$&$+0.00$&$+0.00$&$+0.00$&$+0.00$&$+0.00$&$+0.00$\\
$\hat a_{11}$&$+0.00$&$+0.00$&$+0.00$&$+0.01$&$+0.00$&$+0.00$&$+0.00$&$+0.00$&$+0.00$\\
\hline
$\hat b_{ 3}$&       &$+0.00$&$+0.00$&$+0.00$&$+0.01$&$+0.00$&$+0.01$&$+0.00$&$+0.00$\\
$\hat b_{ 4}$&$+0.02$&       &$+0.01$&$-0.12$&$-0.01$&$+0.00$&$+0.00$&$+0.01$&$+0.01$\\
$\hat b_{ 5}$&$+0.00$&$+0.00$&       &$+0.00$&$+0.00$&$+0.00$&$+0.00$&$+0.00$&$+0.00$\\
$\hat b_{ 6}$&$+0.01$&$+0.01$&$-0.01$&       &$-0.02$&$-0.02$&$-0.01$&$+0.01$&$+0.00$\\
$\hat b_{ 7}$&$+0.00$&$+0.00$&$+0.00$&$+0.00$&       &$+0.00$&$+0.00$&$+0.00$&$+0.00$\\
$\hat b_{ 8}$&$+0.00$&$+0.00$&$+0.00$&$-0.01$&$+0.00$&       &$-0.01$&$-0.01$&$-0.01$\\
$\hat b_{ 9}$&$+0.00$&$+0.00$&$+0.00$&$+0.00$&$+0.00$&$+0.00$&       &$+0.00$&$+0.00$\\
$\hat b_{10}$&$+0.00$&$+0.00$&$+0.00$&$-0.01$&$+0.00$&$-0.02$&$-0.01$&       &$-0.01$\\
$\hat b_{11}$&$+0.00$&$+0.00$&$+0.00$&$+0.00$&$-0.01$&$+0.00$&$+0.00$&$+0.00$&       \\
    \hline
    \end{tabular}
\end{table*}

\begin{figure}
    \centering
    \begin{tikzpicture}
    \node at (0,0) [anchor=north] {\includegraphics[width=0.64\linewidth]{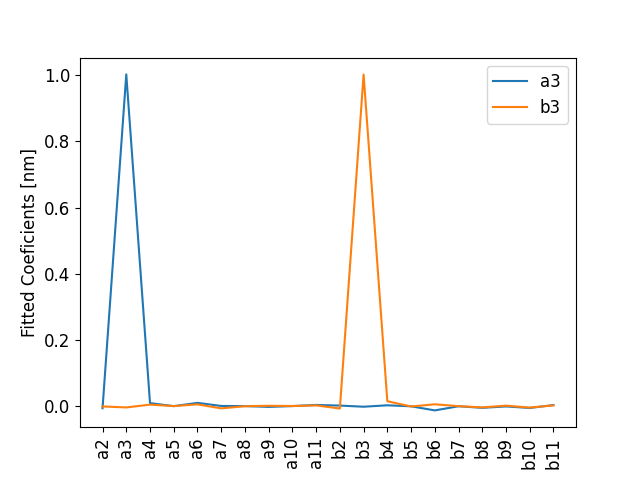}};
    \node at (0,-4.0) [anchor=north] {\includegraphics[width=0.64\linewidth]{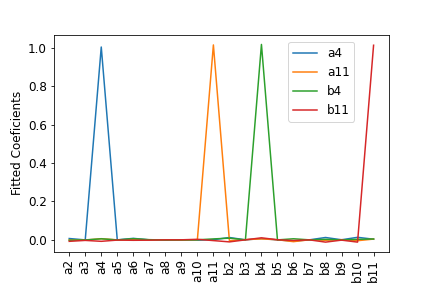}};
    \node at (0,-8.0) [anchor=north] {\includegraphics[width=0.64\linewidth]{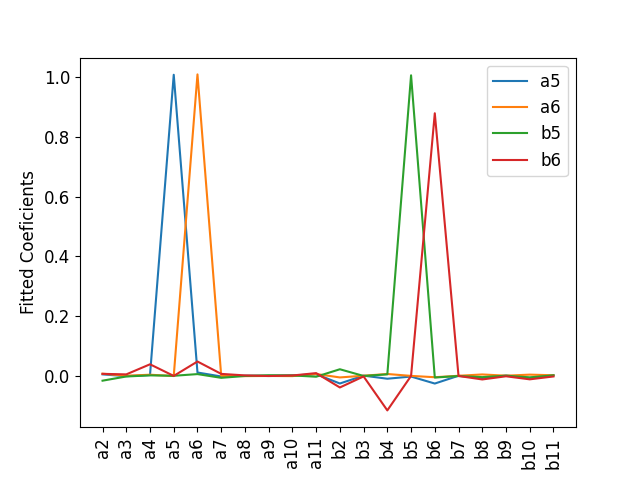}};
    \node at (0,-12.0) [anchor=north] {\includegraphics[width=0.64\linewidth]{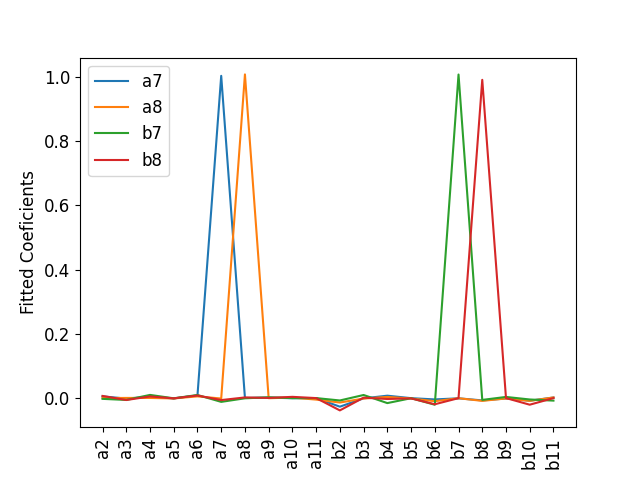}};
    \node at (0,-16.0) [anchor=north] {\includegraphics[width=0.64\linewidth]{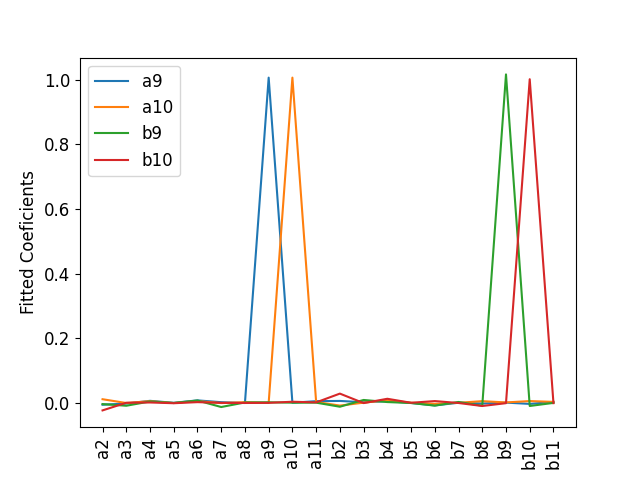}};
    \node at (-3,-0.7) [anchor=west] {(a)};
    \node at (-3,-4.7) [anchor=west] {(b)};
    \node at (-3,-8.7) [anchor=west] {(c)};
    \node at (-3,-12.7) [anchor=west] {(d)};
    \node at (-3,-16.7) [anchor=west] {(e)};
    \end{tikzpicture}
    \caption{Demonstration of the ability of the algorithm to recover single aberrations. The panels show the fitted normalized aberration coefficients when the input simulated images have a single aberration coefficient set to $\lambda/12$. The fitted coefficients have been normalized by dividing by $\lambda/12$. Panel (a) shows the result for tilts $a_3$ and $b_3$. Panel (b) shows defocus and spherical aberration $a_4$, $a_{11}$, $b_4$, and $b_{11}$. Panel (c) shows astigmatisms $a_5$, $a_4$, $b_5$, and $b_6$. Panel (d) shows the comas $a_7$, $a_8$, $b_7$, and $b_8$. Panel (e) shows the trefoils $a_9$, $a_{10}$, $b_9$, and $b_{10}$. We see that for individual aberrations, the algorithm recovers the aberration with little aliasing.}
    \label{figure:test-single}
\end{figure}

We begin by demonstrating the ability of the algorithm to recover single aberrations. For this, we apply the algorithm to simulated data images, consisting of model images with added noise. In this and our other tests, we assume a total signal of about 16,000 electrons, a read noise of 10 electrons, and a background of 40 electrons. The pixel size is $\lambda/3D$ and the slit width is $\lambda/D$ (3 pixels). The controlled displacements are $\pm2\lambda /3D$ ($\pm2$ pixels) and the controlled defocus $db_4$ is $\lambda/7$ RMS. We apply single aberrations of $\lambda/12$ RMS. That is, the simulated data images have all aberration coefficients $a_3$ to $a_{11}$ and $b_3$ to $b_{11}$ equal to zero, except one which is $\lambda/12$. We determined the statistical uncertainties empirically from 10 independent trials.

The results are shown in Table~\ref{table:test-single-a} for $a_3$ to $a_{11}$ and Table~\ref{table:test-single-b} for $b_3$ to $b_{11}$. Each column is the result of one set of simulations. In the first two rows of the body of the tables we show the normalized fitted coefficient $\hat a_j \equiv a_j / (\lambda/12)$ and $\hat b_j \equiv b_j / (\lambda/12)$ corresponding to the coefficient in the simulated data and the empirical uncertainty $\sigma$ in this value. In the lower part of the table we show the mean values of the normalized fitted aberration coefficients that \emph{do not} correspond to the aberration in the simulated data. The uncertainties in these values are typically similar to $\hat\sigma$. The values of $\hat a_j$ and $\hat b_j$ are also shown in Figure~\ref{figure:test-single}.

If the algorithm were ideal, the values of $\hat a_j$ and $\hat b_j$ would be 1 for the applied aberration and 0 otherwise. We see that the actual results are quite close to ideal, with values that should be 1 ranging from 0.879 and 1.017 and values that should be 0 ranging from $-0.12$ to 0.05. It is clear that this slightly non-ideal behaviour is not just the result of noise, since these ranges are larger than expected from the statistical uncertainties. That is, some systematic uncertainty is still present. 

It is notable that the worst case is in $b_6$, astigmatism in the second stage, which shows moderate aliasing with $b_4 $ (defocus in the second stage) at the level of 12\%. At first this puzzled us, as we expected that the aberrations in the second stage would be easy to distinguish in the out-of-focus images. However, when we considered more closely the phase in the second pupil plane, the reason became apparent. Remember that in our simulations the second pupil is twice the size of the first pupil. Thus, the effective phase patterns of both of these aberrations in the second pupil plane are similar, except for a change in sign, because the main differences are in the parts of the second pupil plane that are not illuminated, above and below the illuminated portion. This can be seen by comparing the phase patterns for $b_4$ and $b_6$ in Figure~\ref{figure:catalog-4}. A similar effect can be seen in $a_4$ and $a_6$ in Figure \ref{figure:catalog-1}.

\subsection{Multiple aberrations}

On the strength of these promising results for individual aberrations, we now investigate the ability of the algorithm to recover multiple simultaneous aberrations. We use the same methodology as above. 

In our first example, we have a pixel size of $\lambda/3D$ and a slit width of $\lambda/D$ (three pixels). We apply $a_6 \approx -0.028 \lambda \approx -\lambda/36$ (astigmatism at 45 deg in the first stage), $b_5 \approx -0.025\lambda \approx -\lambda/39$ (astigmatism at 0 deg in the second stage), $b_9 \approx 0.021\lambda \approx \lambda/47$ (trefoil at 30 deg in the second stage), and $b_{11} \approx= 0.044\lambda \approx \lambda/23$ (spherical aberration in the second stage). The total RMS  wave-front error is $0.07\lambda \approx \lambda/14$ and the Strehl ratio is 0.89 (in the absence of the slit). To generate the six images, we apply a shift of $\pm2\lambda/3D$ ($\pm2$ pixels) and a defocus of $db_4 = \lambda/7$.

\begin{figure*}
    \centering
    \begin{tikzpicture}
    \node (a) at (0,0) [anchor=north] {\includegraphics[width=\linewidth]{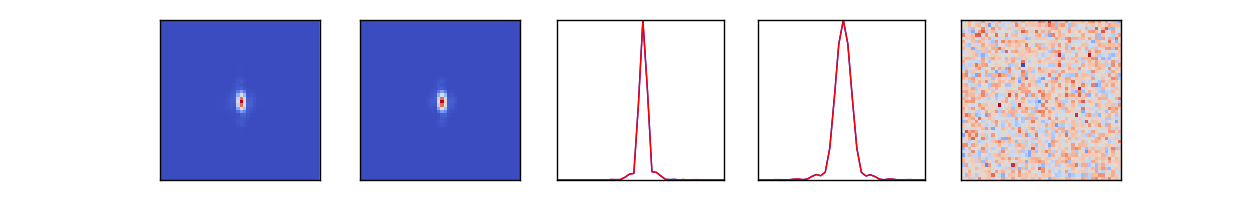}};
    \node (b) at (0,-2.8) [anchor=north] {\includegraphics[width=\linewidth]{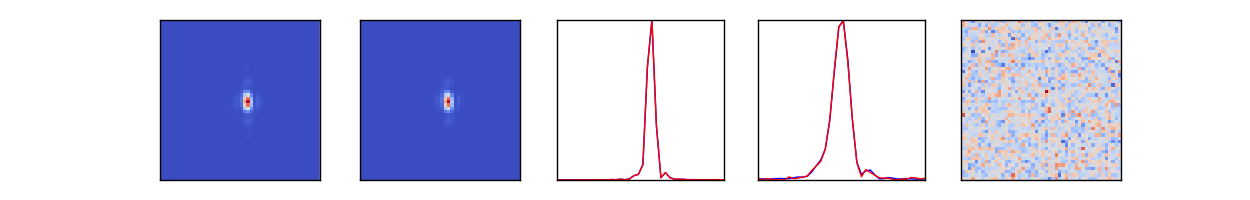}};
    \node (c) at (0,-5.4) [anchor=north] {\includegraphics[width=\linewidth]{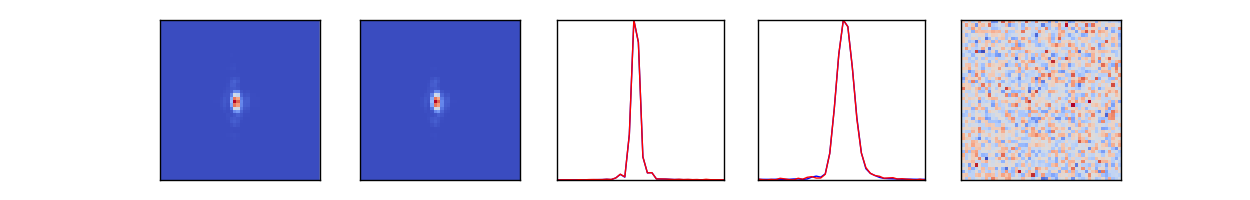}};
    \node (d) at (0,-8.2) [anchor=north] {\includegraphics[width=\linewidth]{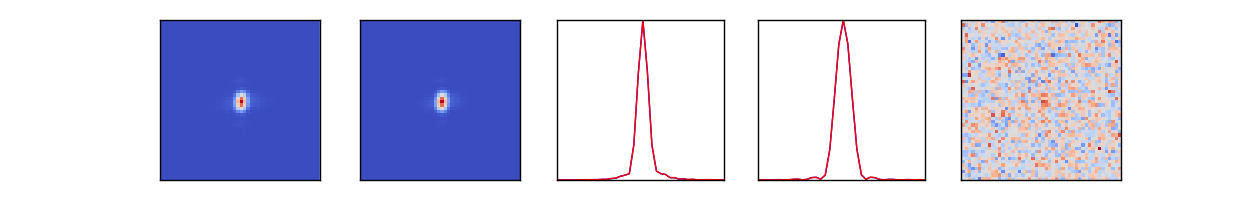}};
    \node (e) at (0,-11.0) [anchor=north] {\includegraphics[width=\linewidth]{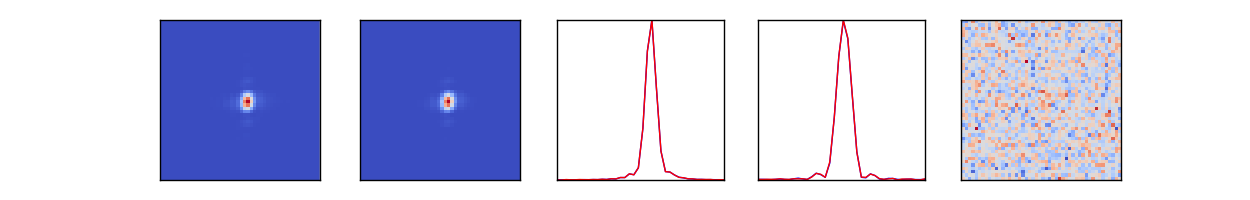}};
    \node (f) at (0,-13.8) [anchor=north] {\includegraphics[width=\linewidth]{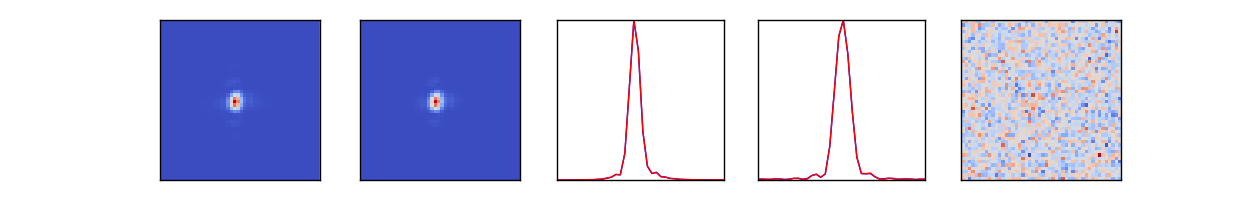}};
    \node at (-5.6,0) [anchor=mid] {Data};
    \node at (-2.7,0) [anchor=mid] {Model};
    \node at (0.2,0) [anchor=mid] {Profiles ($y=0$)};
    \node at (+3.0,0) [anchor=mid] {Profiles ($x=0$)};
    \node at (+5.9,0) [anchor=mid] {Residual};
    \node at (-7.5,-1.50) [anchor=west] {(a)};
    \node at (-7.5,-4.26) [anchor=west] {(b)};
    \node at (-7.5,-7.02) [anchor=west] {(c)};
    \node at (-7.5,-9.78) [anchor=west] {(d)};
    \node at (-7.5,-12.54) [anchor=west] {(e)};
    \node at (-7.5,-15.30) [anchor=west] {(f)};
    \end{tikzpicture}
    \caption{The first multiple-aberration test. Each row corresponds to one of the six images being fitted. The first column shows the image of the simulated data in the exit focal plane, the second column shows the model, the third column shows the overlapping profiles of the data (red) and model (blue) perpendicular to the slit, the fourth column shows the overlapping profiles of the data (red) and model (blue) parallel to the slit, and the final column is the residual.}
    \label{figure:test-multiple-1}
\end{figure*}
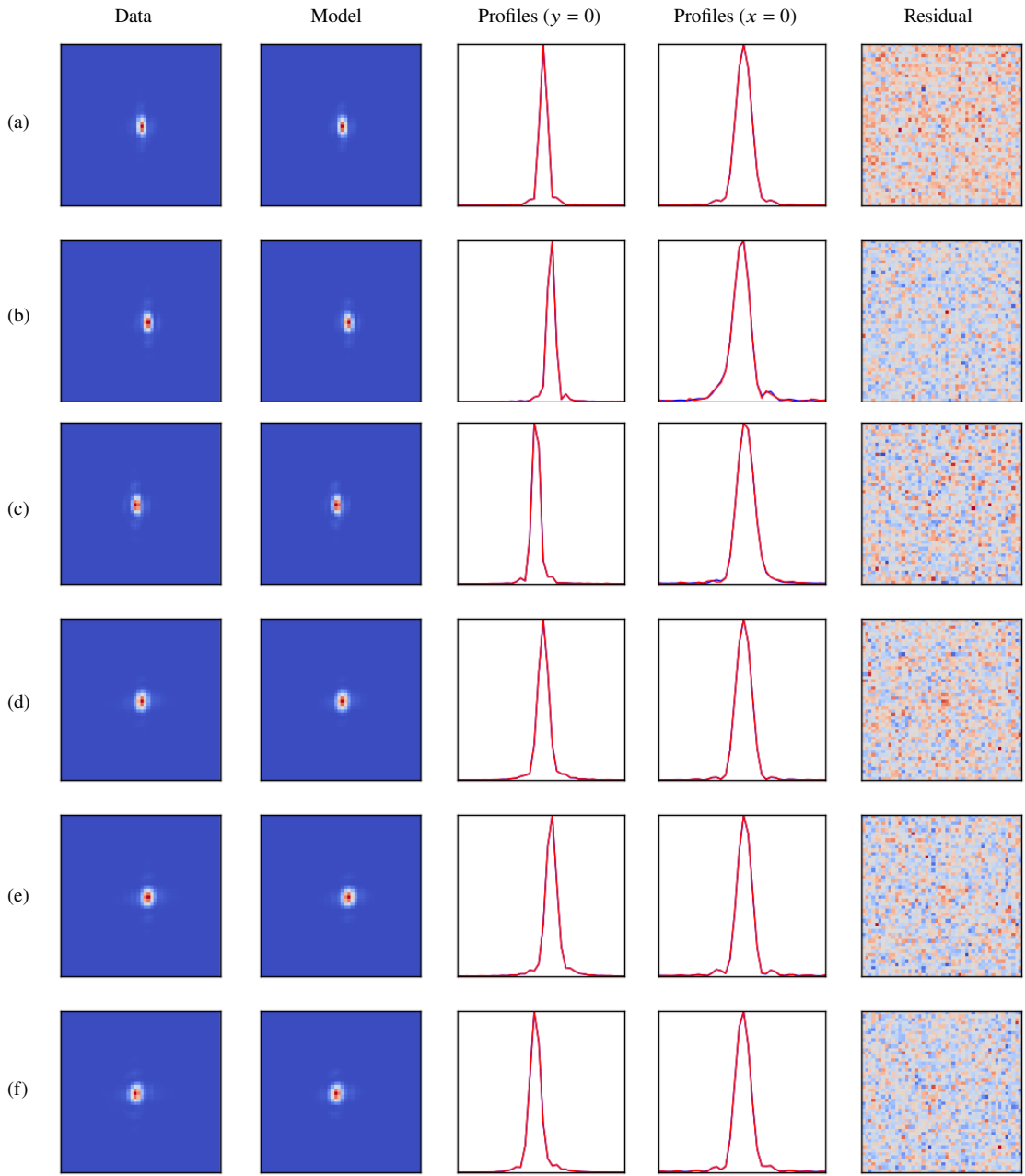

In Figure~\ref{figure:test-multiple-1} we show the results of this test graphically. Each row corresponds to one of the six images being fitted: rows (a) to (c) are in focus and rows (d) to (f) are with the controlled defocus; rows (a) and (d) are with the image centered; rows (b) and (e) are with the slit displaced to the left; and rows (c) and (f) are with the slit displaced to the right. The first column shows the simulated data image in the final focal plane, the second column shows the model, the third column shows the overlapping profiles of the simulated data and model perpendicular to the slit, the fourth column shows the overlapping profiles of the simulated data and model parallel to the slit, and the final column is the residual. As expected, the residuals are close to the noise level. In Table~\ref{table:test-multiple-1} we show the results of this test numerically. We see that the aberrations are recovered with a maximum error of $0.0011\lambda$ (1.6\% of the true RMS wave-front error). Furthermore, no other aberrations were spuriously recovered at more than $0.005\lambda$ (7\%). This suggests that aliasing dominates random errors, but even so it is acceptably weak.

\begin{table*}
    \centering
    \caption{Results of the first multiple-aberration test.}
    \label{table:test-multiple-1}
    \begin{tabular}{lccc}
        \hline
         &Data&Fit&Error  \\
         \hline
         $a_{ 6}$&$-0.02817\lambda$&$-0.02928\lambda \pm 0.00045\lambda$&$+0.00111\lambda \pm 0.00045\lambda$\\
         $b_{ 5}$&$-0.02535\lambda$&$-0.02576\lambda \pm 0.00003\lambda$&$+0.00041\lambda \pm 0.00003\lambda$\\
         $b_{ 9}$&$+0.02113\lambda$&$+0.02158\lambda \pm 0.00001\lambda$&$-0.00045\lambda \pm 0.00001\lambda$\\
         $b_{11}$&$+0.04366\lambda$&$+0.04449\lambda \pm 0.00006\lambda$&$-0.00083\lambda \pm 0.00006\lambda$\\
         \hline
    \end{tabular}
\end{table*}

In our second example, we have a pixel size of $\lambda/3D$ and a slit width of $2\lambda/3D$ (two pixels). We apply  $a_5 \approx -0.025 \lambda \approx -\lambda/39$ (astigmatism at 0 deg in the first stage), $a_{11} \approx 0.028\lambda \approx \lambda/36$ (spherical aberration in the first stage), $b_8 \approx 0.035\lambda \approx \lambda/28$ (coma at 0 deg in the second stage), and $b_9 \approx 0.021\lambda \approx \lambda/47$ (trefoil at 30 deg in the second stage). The total RMS wave-front error is $0.07\lambda \approx \lambda/14$ and the Strehl ratio is 0.91. To generate the six images, we apply a shift of $\pm\lambda/3D$ ($\pm1$ pixel) and a defocus of $db_4 = \lambda/7$.

\begin{figure*}
    \centering
    \begin{tikzpicture}
    \node (a) at (0,0) [anchor=north] {\includegraphics[width=\linewidth]{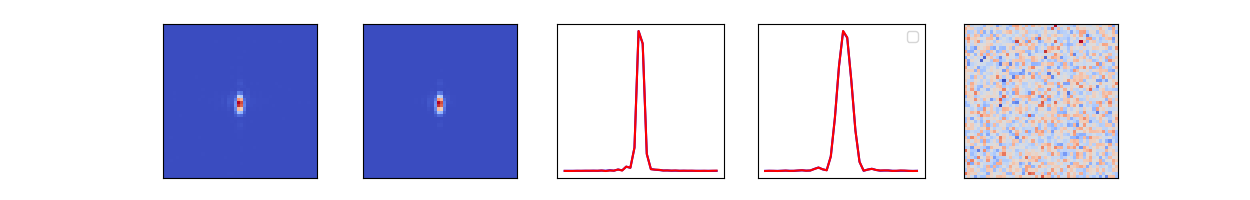}};
    \node (b) at (0,-2.8) [anchor=north] {\includegraphics[width=\linewidth]{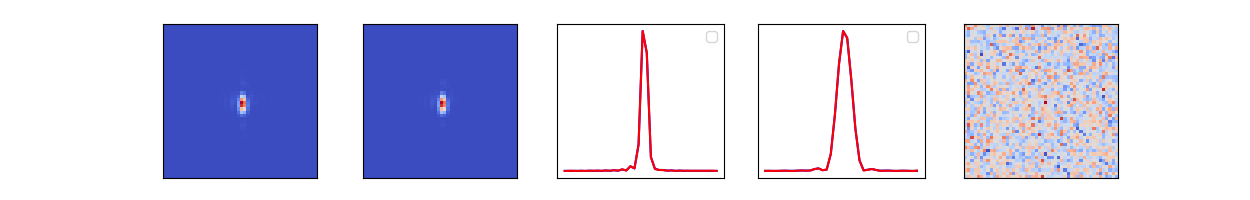}};
    \node (c) at (0,-5.4) [anchor=north] {\includegraphics[width=\linewidth]{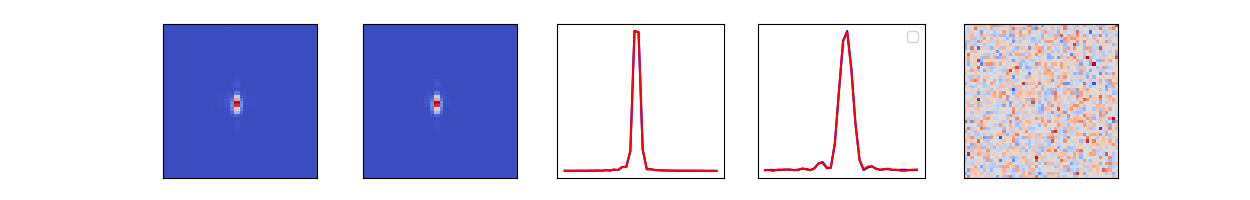}};
    \node (d) at (0,-8.2) [anchor=north] {\includegraphics[width=\linewidth]{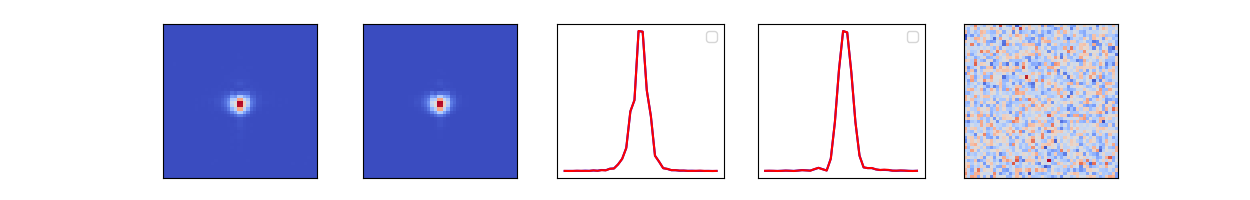}};
    \node (e) at (0,-11.0) [anchor=north] {\includegraphics[width=\linewidth]{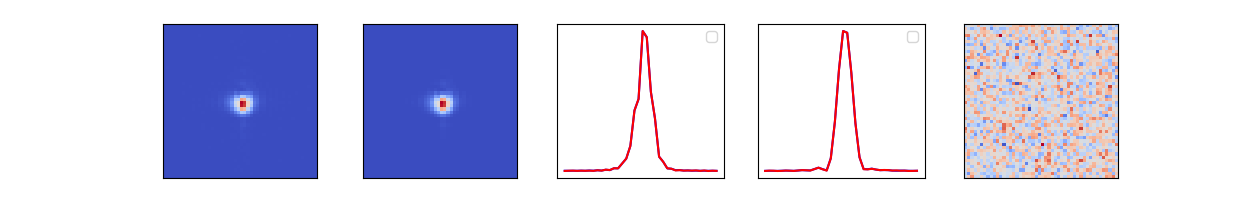}};
    \node (f) at (0,-13.8) [anchor=north] {\includegraphics[width=\linewidth]{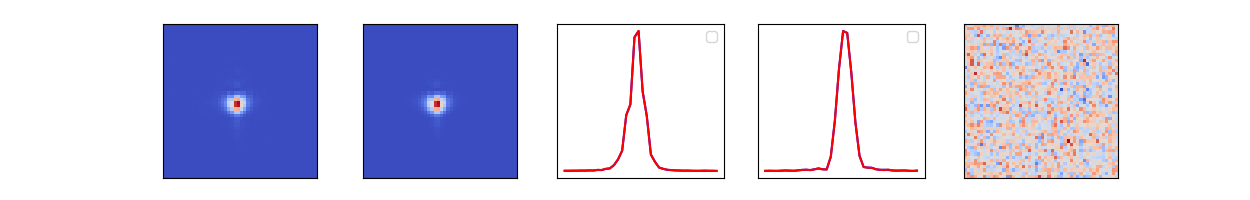}};
    \node at (-5.6,0) [anchor=mid] {Data};
    \node at (-2.7,0) [anchor=mid] {Model};
    \node at (0.2,0) [anchor=mid] {Profiles ($y=0$)};
    \node at (+3.0,0) [anchor=mid] {Profiles ($x=0$)};
    \node at (+5.9,0) [anchor=mid] {Residual};
    \node at (-7.5,-1.50) [anchor=west] {(a)};
    \node at (-7.5,-4.26) [anchor=west] {(b)};
    \node at (-7.5,-7.02) [anchor=west] {(c)};
    \node at (-7.5,-9.78) [anchor=west] {(d)};
    \node at (-7.5,-12.54) [anchor=west] {(e)};
    \node at (-7.5,-15.30) [anchor=west] {(f)};
    \end{tikzpicture}
    \caption{The second multiple-aberration test. Each row corresponds to one of the six images being fitted.  The first column shows the image of the simulated data in the exit focal plane, the second column shows the model, the third column shows the overlapping profiles of the simulated data (red) and model (blue) perpendicular to the slit, the fourth column shows the overlapping profiles of the simulated data (red)  and model (blue) parallel to the slit, and the final column is the residual.}
    \label{figure:test-multiple-2}
\end{figure*}

The results are shown in Figure~\ref{figure:test-multiple-2} and Table~\ref{table:test-multiple-2}. Again, we see that the aberrations are recovered with a maximum error of $0.0033\lambda$ (5\% of the true wave-front error). Furthermore, no other aberrations were spuriously recovered at more than $0.005\lambda$ (7\% of the true wave-front error). Again, aliasing dominates statistical errors, but again it is acceptable.

\begin{table*}
    \centering
    \caption{Results of the second multiple-aberration test.}
    \label{table:test-multiple-2}
    \begin{tabular}{lccc}
        \hline
         &Data&Fit&Error  \\
         \hline
         $a_{ 6}$&$-0.02535\lambda$&$-0.02621\lambda \pm 0.00003\lambda$&$+0.00086\lambda \pm 0.00003\lambda$\\
         $b_{ 5}$&$+0.02817\lambda$&$+0.02928\lambda \pm 0.00004\lambda$&$-0.00111\lambda \pm 0.00004\lambda$\\
         $b_{ 9}$&$+0.03521\lambda$&$+0.03193\lambda \pm 0.00072\lambda$&$+0.00328\lambda \pm 0.00072\lambda$\\
         $b_{11}$&$-0.02113\lambda$&$-0.02159\lambda \pm 0.00003\lambda$&$+0.00046\lambda \pm 0.00003\lambda$\\
         \hline
    \end{tabular}
\end{table*}

We see that in these two tests the algorithm was able to recover the aberrations in the simulated data with excellent precision and without generating significant spurious aberrations. If we correct the aberrations in the data with the fit results, we can improve the image quality. Approximating the Strehl ratio as $S \approx 1 - \sigma^2$ \citep{marechal-1947,born-and-wolf-1975}, in which $\sigma$ is the RMS wave-front error in radians, in the first case we can improve the Strehl ratio from 0.89 to 0.99 and in the second from 0.91 to 0.99.

\section{Future plans}
\label{section:future-plans}

Our future plans are to experimentally verify the method in the laboratory with a simple dual-imaging system with an intermediate slit serving as a stand-in for the spectrograph. Doing so in a laboratory context will allow us to introduce aberrations in a controlled manner and to verify independently the aberrations in each stage.

We also want to consider an optimization to use only four images, three in-focus with controlled displacements perpendicular to the slit and only one out-of-focus to break the degeneracy in the sign of the aberrations. Our motivation for this is that such a sequence of images might place less stringent repeatability requirements on the displacements perpendicular to the slit and on the defocus. That is, one could take an in-focus image displaced one way on the slit, then another in-focus image displaced the other way, then an in-focus image centered on the slit, and finally an out-of-focus images centered on the slit.

\section{Conclusions}
\label{section:conclusions}

We have presented a method for determining the aberrations of a two-stage, long-slit spectrograph working close to the limit of diffraction. We have demonstrated that the algorithm can determine the aberrations in both stages both individually and in combinations with random error of approximately $\lambda/80$ and aliasing typically at a level of about 1\% but sometimes reaching 10\%. We note that our method can be applied to integral-field spectrographs that use image slicers, simply be treating each slice as a slit.

We envisage that our algorithm will be useful during the integration and verification of such spectrographs, as it provides a means to determine the aberrations in the instrument and evaluate corrective measures.

A significant advantage of our method is that data are acquired with the science detector and it does not require hardware beyond a means to feed the spectrograph with a point source that can be displaced perpendicular to the slit and a means to move the detector in focus. This simplicity will often allow determination of the aberrations even after the instrument has been commissioned. One example application would be to verify that the optics have remained aligned during a warm-up and cool-down cycle.

The method could potentially be used to calibrate the non-common-path (NCP) errors between an instrument and the wave-front sensor (WFS) adaptive-optics (AO) system, provided both can be fed with the same point source. The AO system is first commanded to eliminate the aberrations seen by the WFS. Next, our method is used to measure the aberrations seen by the instrument, which are precisely the NCP errors. In this case, the tilt mirror of the AO system can probably be used to command the displacements perpendicular to the slit. (Since our method requires several images, it is not immediately applicable to determining and correction atmospheric aberrations in an AO system.)

We plan to use the algorithm to carry out all of these tasks -- laboratory verification, monitoring, and NCP error calibration –– for the FRIDA spectrograph for GTC.

\section*{Acknowledgements}

We thank Dr.\ Jorge Fuentes-Fernández and an anonymous referee for useful comments.

%%%%%%%%%%%%%%%%%%%%%%%%%%%%%%%%%%%%%%%%%%%%%%%%%%
\section*{Data Availability}

The data underlying this article will be shared on reasonable request to the corresponding author.



%\begin{thebibliography}{99}
%\end{thebibliography}

%%%%%%%%%%%%%%%%%%%%%%%%%%%%%%%%%%%%%%%%%%%%%%%%%%

%%%%%%%%%%%%%%%%% APPENDICES %%%%%%%%%%%%%%%%%%%%%

\clearpage
\appendix

\section{Catalog of Aberrated Images}
\label{appendix:catalog}

In this appendix we present images of a point source imaged through a two-stage spectrograph in the presence of small aberrations and different projections on the slit. The pixel size is $\lambda/3D$, the slit width is $\lambda/D$ (3 pixels), and the displacements are $\pm\lambda/3D$ ($\pm1$ pixel).

\begin{figure*}
    \centering
    \begin{tikzpicture}
    \node (a) at (0,0) [anchor=north] {\includegraphics[width=0.7\linewidth]{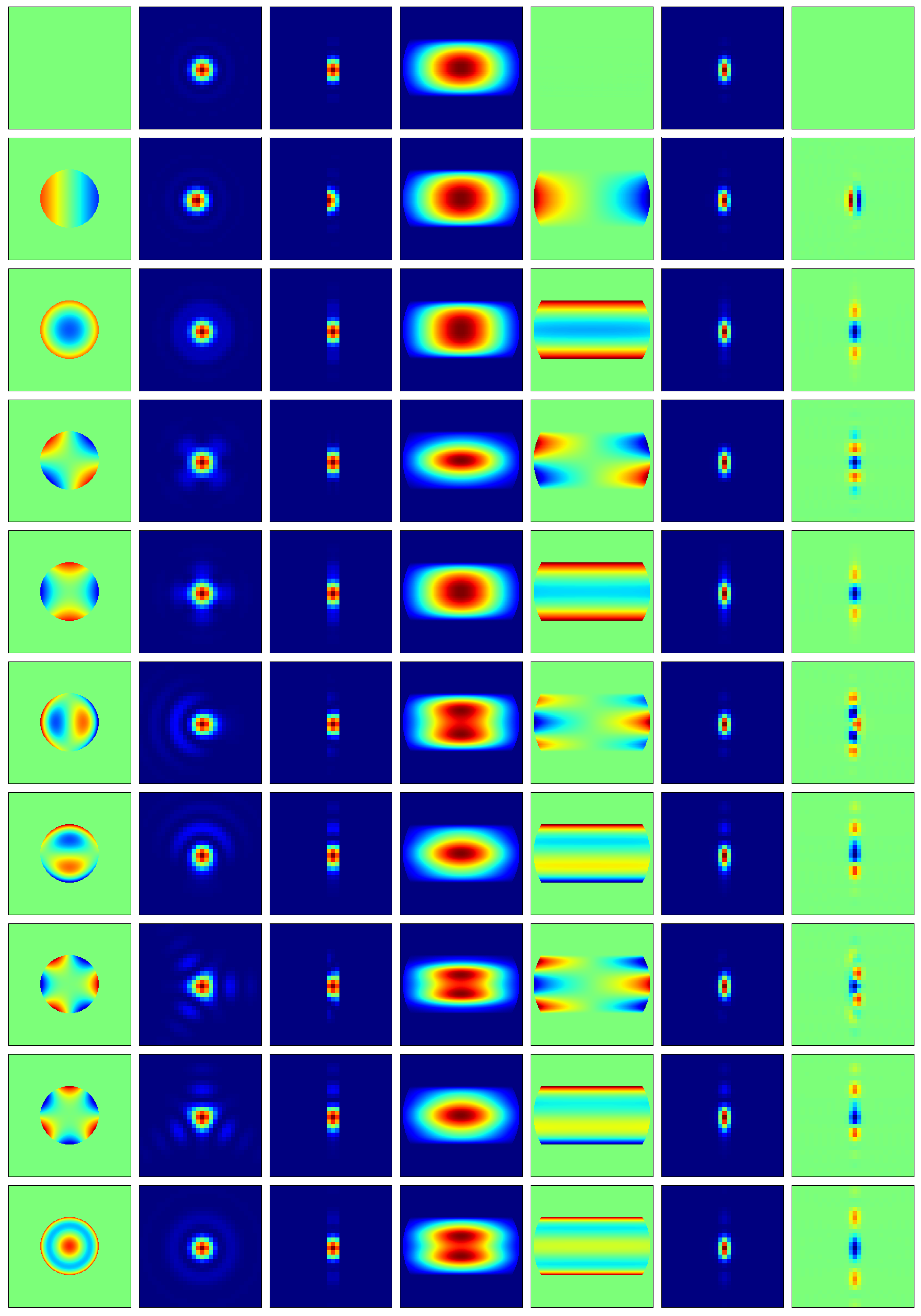}};
    \node at (-5.3,0.2) [anchor=mid] {$\phi_1$};
    \node at (-3.5,0.2) [anchor=mid] {$I_1$};
    \node at (-1.7,0.2) [anchor=mid] {$SI_1$};
    \node at (-0.1,0.2) [anchor=mid] {$A_2$};
    \node at (+1.7,0.2) [anchor=mid] {$\phi_2$};
    \node at (+3.6,0.2) [anchor=mid] {$I_2$};
    \node at (+5.3,0.2) [anchor=mid] {$\Delta I_2$};
    \node at (-8.5,-1) [anchor=west] {(a)};
    \node at (-8.5,-2.78) [anchor=west] {(b) $a_3 = \lambda/12$};
    \node at (-8.5,-4.56) [anchor=west] {(c) $a_4 = \lambda/12$};
    \node at (-8.5,-6.33) [anchor=west] {(d) $a_5 = \lambda/12$};
    \node at (-8.5,-8.11) [anchor=west] {(e) $a_6 = \lambda/12$};
    \node at (-8.5,-9.89) [anchor=west] {(f) $a_7 = \lambda/12$};
    \node at (-8.5,-11.67) [anchor=west] {(g) $a_8 = \lambda/12$};
    \node at (-8.5,-13.44) [anchor=west] {(h) $a_9 = \lambda/12$};
    \node at (-8.5,-15.22) [anchor=west] {(i) $a_{10} = \lambda/12$};
    \node at (-8.5,-17.00) [anchor=west] {(j) $a_{11} = \lambda/12$};
    \end{tikzpicture}
    \caption{Images showing a point source imaged through the spectrograph. The pixel size is $\lambda/3D$ and the slit width is $\lambda/D$ (3 pixels). The point source is centered on the slit. The top row is without aberrations. The subsequent rows have $a_3$ to $a_{11}$ in turn set to $\lambda/12$ RMS. See the text for a description of the columns.}
    \label{figure:catalog-1}
\end{figure*}

\begin{figure*}
    \centering
    \begin{tikzpicture}
    \node (a) at (0,0) [anchor=north] {\includegraphics[width=0.7\linewidth]{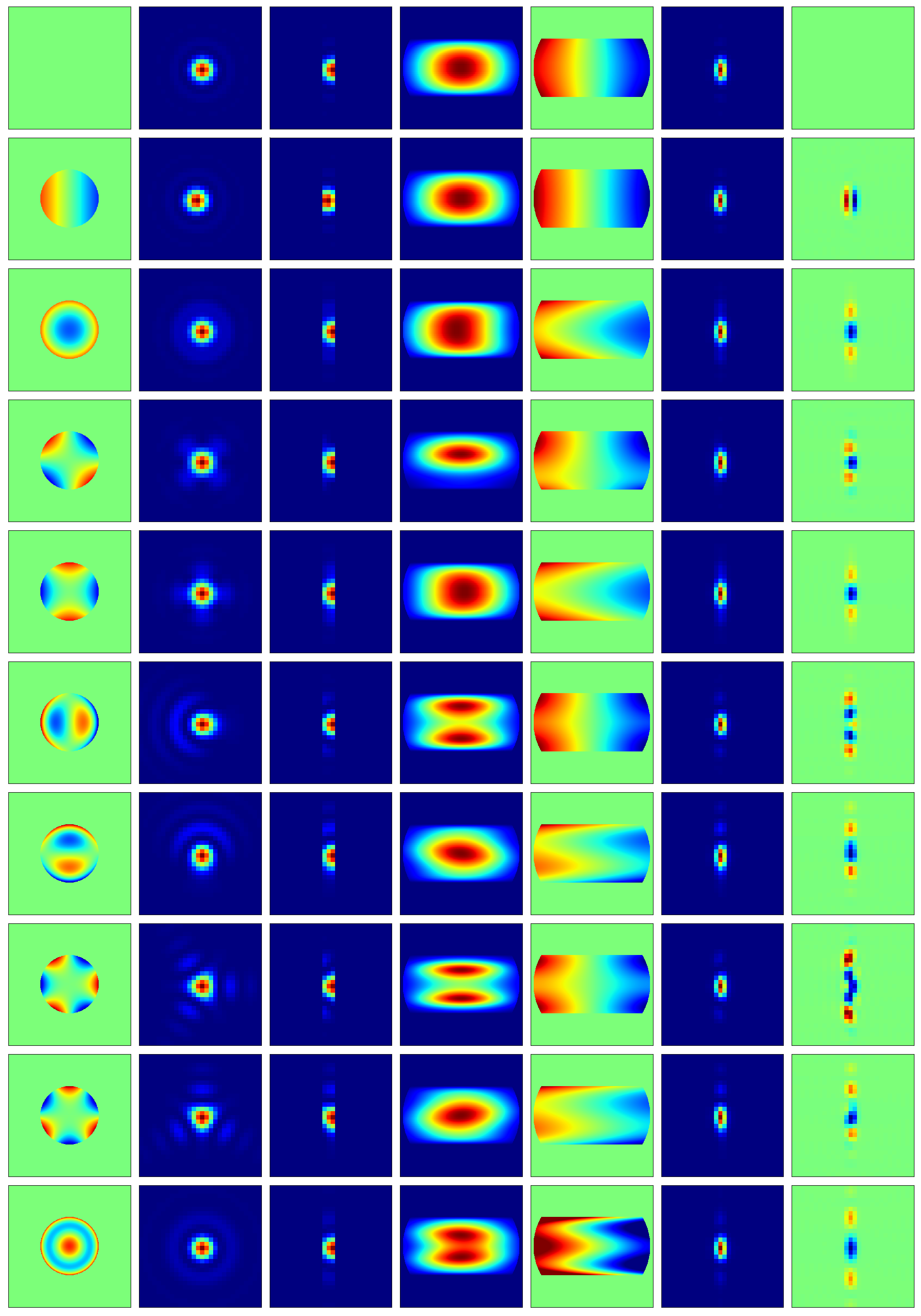}};
    \node at (-5.3,0.2) [anchor=mid] {$\phi_1$};
    \node at (-3.5,0.2) [anchor=mid] {$I_1$};
    \node at (-1.7,0.2) [anchor=mid] {$SI_1$};
    \node at (-0.1,0.2) [anchor=mid] {$A_2$};
    \node at (+1.7,0.2) [anchor=mid] {$\phi_2$};
    \node at (+3.6,0.2) [anchor=mid] {$I_2$};
    \node at (+5.3,0.2) [anchor=mid] {$\Delta I_2$};
    \node at (-8.5,-1) [anchor=west] {(a)};
    \node at (-8.5,-2.78) [anchor=west] {(b) $a_3 = \lambda/12$};
    \node at (-8.5,-4.56) [anchor=west] {(c) $a_4 = \lambda/12$};
    \node at (-8.5,-6.33) [anchor=west] {(d) $a_5 = \lambda/12$};
    \node at (-8.5,-8.11) [anchor=west] {(e) $a_6 = \lambda/12$};
    \node at (-8.5,-9.89) [anchor=west] {(f) $a_7 = \lambda/12$};
    \node at (-8.5,-11.67) [anchor=west] {(g) $a_8 = \lambda/12$};
    \node at (-8.5,-13.44) [anchor=west] {(h) $a_9 = \lambda/12$};
    \node at (-8.5,-15.22) [anchor=west] {(i) $a_{10} = \lambda/12$};
    \node at (-8.5,-17.00) [anchor=west] {(j) $a_{11} = \lambda/12$};
    \end{tikzpicture}
    \caption{Images showing a point source imaged through the spectrograph. The pixel size is $\lambda/3D$ and the slit width is $\lambda/D$ (3 pixels). The point source is displaced by $\lambda/3D$ (1 pixel) to the left. The top row is without aberrations. The subsequent rows have $a_3$ to $a_{11}$ in turn set to  $\lambda/12$ RMS. See the text for a description of the columns.}
    \label{figure:catalog-2}
\end{figure*}

\begin{figure*}
    \centering
    \begin{tikzpicture}
    \node (a) at (0,0) [anchor=north] {\includegraphics[width=0.7\linewidth]{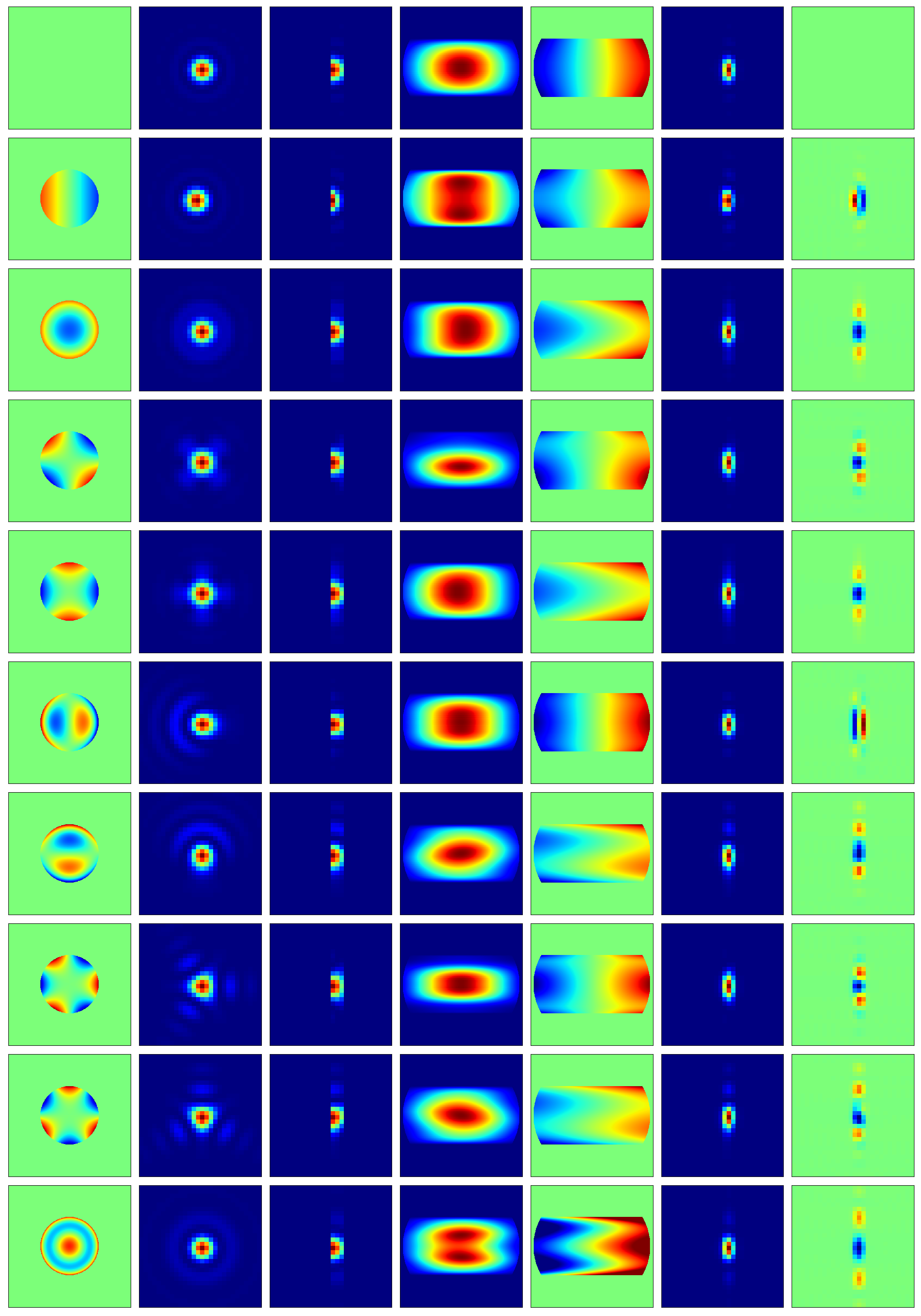}};
    \node at (-5.3,0.2) [anchor=mid] {$\phi_1$};
    \node at (-3.5,0.2) [anchor=mid] {$I_1$};
    \node at (-1.7,0.2) [anchor=mid] {$SI_1$};
    \node at (-0.1,0.2) [anchor=mid] {$A_2$};
    \node at (+1.7,0.2) [anchor=mid] {$\phi_2$};
    \node at (+3.6,0.2) [anchor=mid] {$I_2$};
    \node at (+5.3,0.2) [anchor=mid] {$\Delta I_2$};
    \node at (-8.5,-1) [anchor=west] {(a)};
    \node at (-8.5,-2.78) [anchor=west] {(b) $a_3 = \lambda/12$};
    \node at (-8.5,-4.56) [anchor=west] {(c) $a_4 = \lambda/12$};
    \node at (-8.5,-6.33) [anchor=west] {(d) $a_5 = \lambda/12$};
    \node at (-8.5,-8.11) [anchor=west] {(e) $a_6 = \lambda/12$};
    \node at (-8.5,-9.89) [anchor=west] {(f) $a_7 = \lambda/12$};
    \node at (-8.5,-11.67) [anchor=west] {(g) $a_8 = \lambda/12$};
    \node at (-8.5,-13.44) [anchor=west] {(h) $a_9 = \lambda/12$};
    \node at (-8.5,-15.22) [anchor=west] {(i) $a_{10} = \lambda/12$};
    \node at (-8.5,-17.00) [anchor=west] {(j) $a_{11} = \lambda/12$};
    \end{tikzpicture}
    \caption{Images showing a point source imaged through the spectrograph. The pixel size is $\lambda/3D$ and the slit width is $\lambda/D$ (3 pixels). The point source is displaced by $\lambda/3D$ (1 pixel) to the right. The top row is without aberrations. The subsequent rows have $a_3$ to $a_{11}$ in turn set to  $\lambda/12$ RMS. See the text for a description of the columns.}
    \label{figure:catalog-3}
\end{figure*}

\begin{figure*}
    \centering
    \begin{tikzpicture}
    \node (a) at (0,0) [anchor=north] {\includegraphics[width=0.7\linewidth]{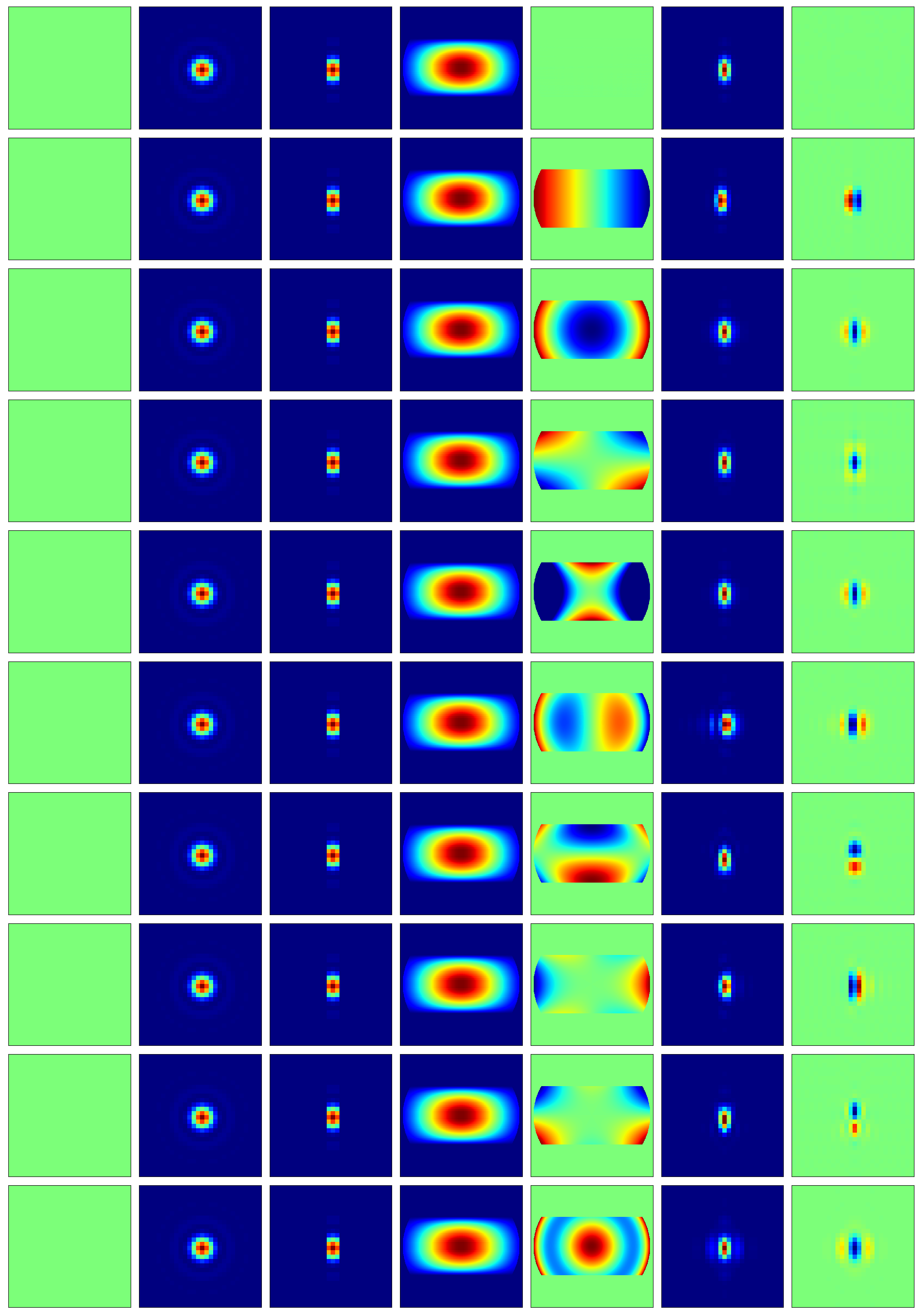}};
    \node at (-5.3,0.2) [anchor=mid] {$\phi_1$};
    \node at (-3.5,0.2) [anchor=mid] {$I_1$};
    \node at (-1.7,0.2) [anchor=mid] {$SI_1$};
    \node at (-0.1,0.2) [anchor=mid] {$A_2$};
    \node at (+1.7,0.2) [anchor=mid] {$\phi_2$};
    \node at (+3.6,0.2) [anchor=mid] {$I_2$};
    \node at (+5.3,0.2) [anchor=mid] {$\Delta I_2$};
    \node at (-8.5,-1) [anchor=west] {(a)};
    \node at (-8.5,-2.78) [anchor=west] {(b) $b_3 = \lambda/12$};
    \node at (-8.5,-4.56) [anchor=west] {(c) $b_4 = \lambda/12$};
    \node at (-8.5,-6.33) [anchor=west] {(d) $b_5 = \lambda/12$};
    \node at (-8.5,-8.11) [anchor=west] {(e) $b_6 = \lambda/12$};
    \node at (-8.5,-9.89) [anchor=west] {(f) $b_7 = \lambda/12$};
    \node at (-8.5,-11.67) [anchor=west] {(g) $b_8 = \lambda/12$};
    \node at (-8.5,-13.44) [anchor=west] {(h) $b_9 = \lambda/12$};
    \node at (-8.5,-15.22) [anchor=west] {(i) $b_{10} = \lambda/12$};
    \node at (-8.5,-17.00) [anchor=west] {(j) $b_{11} = \lambda/12$};
    \end{tikzpicture}
    \caption{Images showing a point source imaged through the spectrograph. The pixel size is $\lambda/3D$ and the slit width is $\lambda/D$ (3 pixels). The point source is centered on the slit. The top row is without aberrations. The subsequent rows have $b_3$ to $b_{11}$ in turn set to  $\lambda/12$ RMS. See the text for a description of the columns.}
    \label{figure:catalog-4}
\end{figure*}

\begin{figure*}
    \centering
    \begin{tikzpicture}
    \node (a) at (0,0) [anchor=north] {\includegraphics[width=0.7\linewidth]{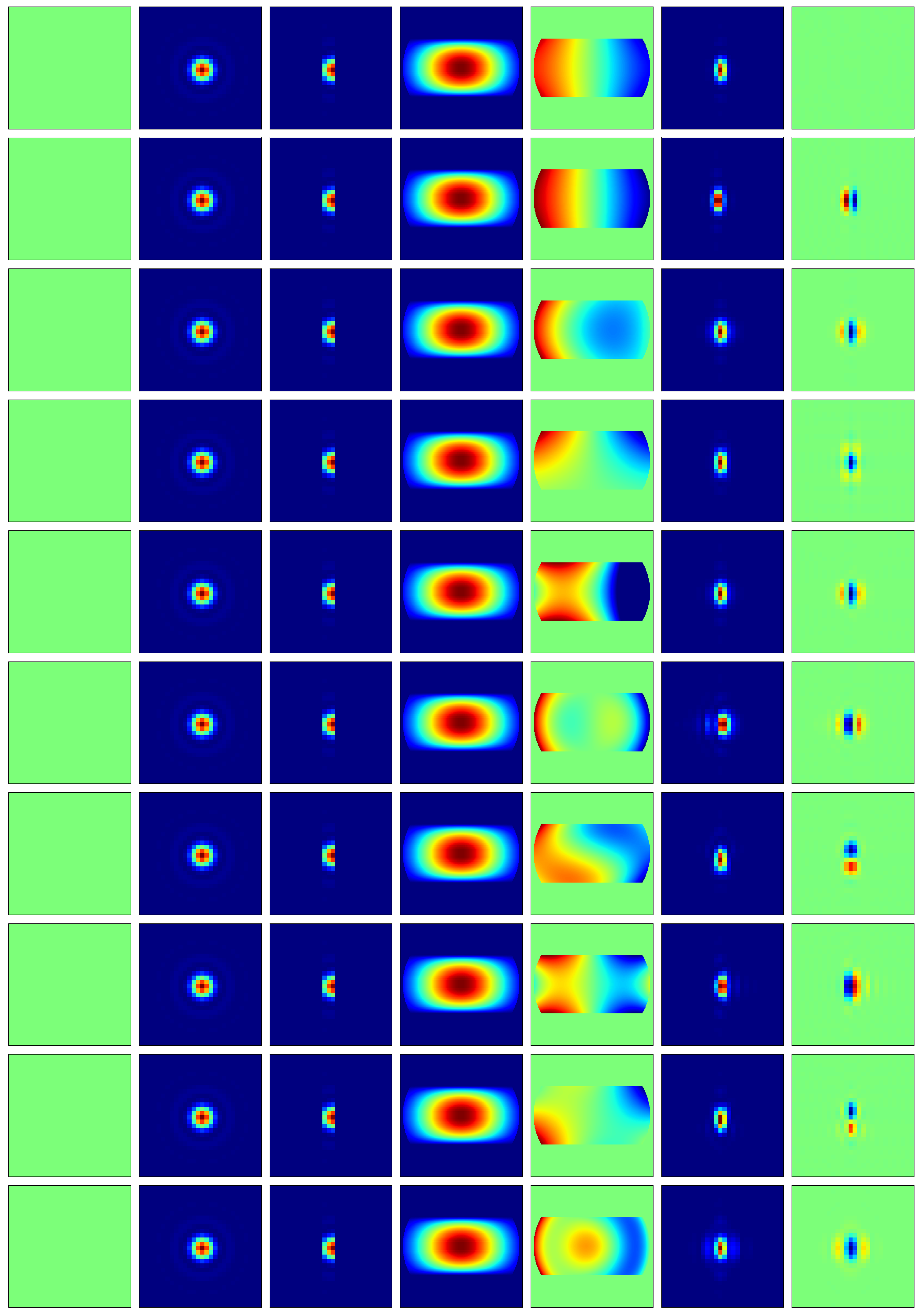}};
    \node at (-5.3,0.2) [anchor=mid] {$\phi_1$};
    \node at (-3.5,0.2) [anchor=mid] {$I_1$};
    \node at (-1.7,0.2) [anchor=mid] {$SI_1$};
    \node at (-0.1,0.2) [anchor=mid] {$A_2$};
    \node at (+1.7,0.2) [anchor=mid] {$\phi_2$};
    \node at (+3.6,0.2) [anchor=mid] {$I_2$};
    \node at (+5.3,0.2) [anchor=mid] {$\Delta I_2$};
    \node at (-8.5,-1) [anchor=west] {(a)};
    \node at (-8.5,-2.78) [anchor=west] {(b) $b_3 = \lambda/12$};
    \node at (-8.5,-4.56) [anchor=west] {(c) $b_4 = \lambda/12$};
    \node at (-8.5,-6.33) [anchor=west] {(d) $b_5 = \lambda/12$};
    \node at (-8.5,-8.11) [anchor=west] {(e) $b_6 = \lambda/12$};
    \node at (-8.5,-9.89) [anchor=west] {(f) $b_7 = \lambda/12$};
    \node at (-8.5,-11.67) [anchor=west] {(g) $b_8 = \lambda/12$};
    \node at (-8.5,-13.44) [anchor=west] {(h) $b_9 = \lambda/12$};
    \node at (-8.5,-15.22) [anchor=west] {(i) $b_{10} = \lambda/12$};
    \node at (-8.5,-17.00) [anchor=west] {(j) $b_{11} = \lambda/12$};
    \end{tikzpicture}
    \caption{Images showing a point source imaged through the spectrograph. The pixel size is $\lambda/3D$ and the slit width is $\lambda/D$ (3 pixels). The point source is displaced by $\lambda/3D$ (1 pixel) to the left. The top row is without aberrations. The subsequent rows have $b_3$ to $b_{11}$ in turn set to  $\lambda/12$ RMS. See the text for a description of the columns.}
    \label{figure:catalog-5}
\end{figure*}

\begin{figure*}
    \centering
    \begin{tikzpicture}
    \node (a) at (0,0) [anchor=north] {\includegraphics[width=0.7\linewidth]{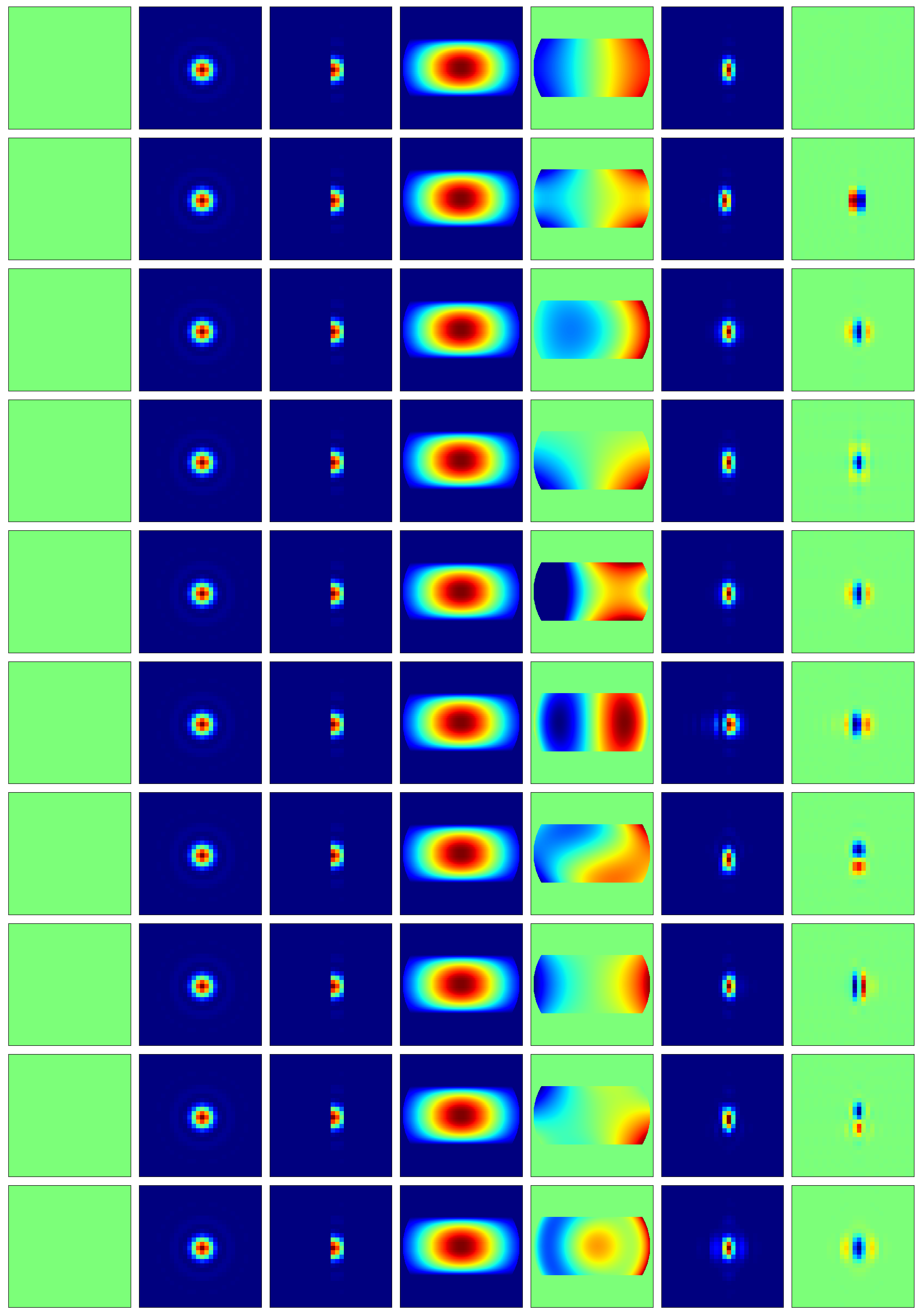}};
    \node at (-5.3,0.2) [anchor=mid] {$\phi_1$};
    \node at (-3.5,0.2) [anchor=mid] {$I_1$};
    \node at (-1.7,0.2) [anchor=mid] {$SI_1$};
    \node at (-0.1,0.2) [anchor=mid] {$A_2$};
    \node at (+1.7,0.2) [anchor=mid] {$\phi_2$};
    \node at (+3.6,0.2) [anchor=mid] {$I_2$};
    \node at (+5.3,0.2) [anchor=mid] {$\Delta I_2$};
    \node at (-8.5,-1) [anchor=west] {(a)};
    \node at (-8.5,-2.78) [anchor=west] {(b) $b_3 = \lambda/12$};
    \node at (-8.5,-4.56) [anchor=west] {(c) $b_4 = \lambda/12$};
    \node at (-8.5,-6.33) [anchor=west] {(d) $b_5 = \lambda/12$};
    \node at (-8.5,-8.11) [anchor=west] {(e) $b_6 = \lambda/12$};
    \node at (-8.5,-9.89) [anchor=west] {(f) $b_7 = \lambda/12$};
    \node at (-8.5,-11.67) [anchor=west] {(g) $b_8 = \lambda/12$};
    \node at (-8.5,-13.44) [anchor=west] {(h) $b_9 = \lambda/12$};
    \node at (-8.5,-15.22) [anchor=west] {(i) $b_{10} = \lambda/12$};
    \node at (-8.5,-17.00) [anchor=west] {(j) $b_{11} = \lambda/12$};
    \end{tikzpicture}
    \caption{Images showing a point source imaged through the spectrograph. The pixel size is $\lambda/3D$ and the slit width is $\lambda/D$ (3 pixels). The point source is displaced by $\lambda/3D$ (1 pixel) to the right. The top row is without aberrations. The subsequent rows have $b_3$ to $b_{11}$ in turn set to  $\lambda/12$ RMS. See the text for a description of the columns.}
    \label{figure:catalog-6}
\end{figure*}

Figures~\ref{figure:catalog-1} to \ref{figure:catalog-3} show the effects of aberrations in the first stage, $a_3$ to $a_{11}$, and Figures~\ref{figure:catalog-4} to \ref{figure:catalog-6} show the effects of aberrations in the second stage, $b_3$ to $b_{11}$. In all case, the amplitude of the applied aberration is $\lambda/12$ RMS.

In each figure, the columns show, from left to right: the phase $\phi_1$  in the pupil plane of the first stage; the image $I_1$ in the exit focal plane of the first stage just before the slit; the image $SI_1$  in the exit focal plane of the first stage just after the slit (which is vertical); the amplitude $A_2$ of the intensity in the pupil plane of the second stage; the phase  $\phi_2$ of the intensity in the pupil plane of the second stage; the image $I_2$   in the exit focal plane of the second stage; and finally the difference $\Delta I_2$ between the aberrated image and the unaberrated image.

% Don't change these lines
\bsp	% typesetting comment
\label{lastpage}
\end{document}